\documentstyle[aps,pra,prabib,psfig,epsfig]{revtex}

\renewcommand{\narrowtext}{\widetext}

\begin{document}

\title{QED self-energy contribution to highly-excited atomic states}

\author{\' Eric-Olivier
  Le~Bigot\thanks{\texttt{lebigot@spectro.jussieu.fr}}
\and Paul Indelicato\thanks{\texttt{paul@spectro.jussieu.fr}}
}
\address{
Laboratoire Kastler-Brossel,
\' Ecole Normale Sup\' erieure et Universit\' e P. et M. Curie\\
Unit\' e Mixte de Recherche du CNRS n$^\circ$ C8552\\
Case 74; 4, pl.~Jussieu, 75252 Paris CEDEX 05, France}
\author{Peter J. Mohr\thanks{\texttt{mohr@nist.gov}}}
\address{
  National Institute of Standards and Technology,
  Gaithersburg, Maryland 20899-8401
 }

\date{\today}

\maketitle

\begin{abstract}
We present numerical values for the
  self-energy shifts predicted by QED (Quantum Electrodynamics) for
  hydrogenlike ions (nuclear charge $60 \le Z \le 110$) with an
  electron in an $n=3$, $4$ or $5$ level with high angular momentum ($5/2
  \le j \le 9/2$). 
  Applications include predictions of precision transition energies and
  studies of the outer-shell
  structure of atoms and ions.
\end{abstract}

\narrowtext

\section{Introduction}

The
one-loop self-energy is the largest radiative correction in atoms and
ions. It has been known for many years that for high nuclear charge
$Z$, results
obtained by a perturbation expansion in the number of interactions
with the nucleus, i.e., in powers of $Z\alpha$ are not accurate. There
are many recent examples in which self-energy shifts for high
principal quantum numbers and/or angular momenta are needed. For
example, the outer-shell structure of very heavy elements is being
studied to determine electron affinities or chemical properties.
Similarly, transitions within the ground configuration of the Ti-like
iso-electronic sequence (with outer-shell structure $3d^4$) have been
measured for very high nuclear charges
$Z$ \cite{mstm95,smmg96,sbmg97,tbuc98,cbwd99,ubb2000,wcc2001}, with results
that differ systematically from theoretical
calculations \cite{fsi90,pmi94,bec97,bec99} that
do not include the self-energy correction for the $3d_{5/2}$ level beyond
the Bethe logarithm and lowest-order electron anomalous moment.  The
same is true for systematic studies of magnesiumlike
ions \cite{skir89}. Yet it has been shown recently that effects beyond
the lowest order are large at high $Z$, even for rather large $n$ and
angular momenta~\cite{iam98a}.  Several calculations have thus been
undertaken.  Preliminary results have been reported for the $6s$ and
$8d_{3/2}$ shells~\cite{iam98a}, while Labzowski \emph{et
  al.}~\cite{lgtp99} and Yerokhin and Shabaev \cite{yas99} have
calculated
a limited number of self-energy level shifts with methods developed
recently. In this paper, we provide accurate calculations for
nuclear charge $Z$ in the range
$60 \le Z \le 110$ for an electron in a level with $n=3$, $4$, or $5$ 
with high angular momentum ($5/2 \le j \le 9/2$).  The results
increase considerably
the number of atoms for which the exact one-loop self-energy is
available.

After the seminal works of Brown \emph{et al.}~\cite{bls59} and
of Desiderio and Johnson \cite{daj71}, one of us (PJM)
developed an efficient method for evaluating the self-energy
  level shifts in a framework where the electron-nucleus interaction
  is treated non-perturbatively. This method was first applied to the
$1s$ level of one-electron atoms \cite{moh74a,moh74b}. Later,
  the self-energy contribution to the Lamb shift of the $2s$ and
  $2p_{1/2}$ states was studied~\cite{moh75},
and subsequently, all levels with principal quantum number $n=2$ were
evaluated with higher accuracy \cite{moh82}. Additional excited states with
$3\le n\le 5$ and angular momentum $j \le 3/2$ were studied
more recently in Ref.~\cite{mak92}, where the difficulties which arise
for highly-excited states are described and
solved for low angular momenta. Self-energy calculations for states
with angular momentum $j>3/2$ has been a long standing problem since the
publication of the original method~\cite{moh74a}.

In the present paper we extend the self-energy calculation methods
developed in \cite{moh74a,moh74b,moh75,moh82,mak92} to \emph{arbitrary}
angular momenta. We derive general, numerically efficient formulas
for the angular integrals that appear in self-energy expressions;
these integrals were only known for states with $j\le 3/2$. We perform
calculations for $5/2 \le j\le 9/2$ and for principal
quantum numbers $n=3$, $4$  and $5$. 
The numerically efficient renormalization technique described in
Refs.~\cite{iam92,iam98} is used, as its formulation is independent of
the atomic level. 
This method yields precise results for the
self-energy shifts; other groups have published numerical results
based on various numerical strategies that give less precise
values~\cite{v99,qag94,pls93}.

The implementation of our formulas on a computer is simple,
  since the only mathematical objects used in our final expressions
  are Bessel functions of the first kind and (squared) Wigner $3j$
  symbols with all angular momentum projections set to zero. Our
expressions are optimized for numerical calculations and have proved
to yield very accurate results.  Our formulas for electrons with
angular momentum $j=3/2$ differ from known
expressions~\cite{moh82,moh92}, because we have adapted our results to
more accurate numerical calculations.

The outline of the paper is as follows. In Sec.~\ref{sec:princ}, we recall the
principle of the self-energy calculation we use.  In Sec.~\ref{sec:form-angul}
we summarize our formulas for the self-energy angular integrals, which
constitute the main analytical result of the present work. In
Sec.~\ref{sec:numer-res} we present numerical results for levels 
$d_{5/2}$,
$f_{5/2}$, $f_{7/2}$, $g_{7/2}$ and $g_{9/2}$ 
with principal quantum
numbers $3 \le n \le 5$.
Detailed derivations of the formulas presented in
Sec.~\ref{sec:form-angul} are given in Sec.~\ref{sec:optnumder}, while
derivations by
a different method, used for checking purposes, are presented
in Sec.~\ref{subsec:direct}. Section~\ref{sec:concl} concludes the paper.

Throughout this article, we use the convention $\sum_{i=a}^b \equiv 0$
whenever $a>b$.

\section{Self-energy shift formula}
\label{sec:princ}

The expression for the self-energy shift of an electronic state $n$
can be written for a large class of potentials ${\cal V}(\bbox{x})$ as
the sum
\begin{equation}
\label{eq:defSE}
{\cal E}_{\rm SE} \equiv {\cal E}_{\rm L} + {\cal E}_{\rm H}
\end{equation}
of a low-energy part $ {\cal E}_{\rm L}$ and a high-energy part 
$ {\cal E}_{\rm H}$ given (in units in which $\hbar=c=m_{\rm e} = 1)$
by \cite{moh74a}
\widetext
\begin{equation} 
 {\cal E}_{\rm L}\equiv\frac{\alpha} {\pi} {\cal E}_n - \frac{\alpha} {\pi}
{\mathrm P}\!\int_0^{{\cal E}_n}
d z 
\int d\bbox x_2 \int  d\bbox x_1 \, \varphi_n^\dagger (\bbox x_2) \alpha^l
{\cal G}(\bbox x_2,\bbox x_1, z) \alpha^m \varphi_n(\bbox x_1) (\delta_{l,m}\bbox
\nabla_2\cdot \bbox \nabla_1 - \nabla_2^l \nabla_1^m) \frac{\sin 
[({\cal E}_n-z)x_{21}] }{({\cal E}_n-z)^2 x_{21}}
\label{eq:lowe} \end{equation} 
\noindent and
\begin{equation} 
\  {\cal E}_{\rm H}\equiv \frac{\alpha}{ 2\pi {i}} \int_{\rm C_H} d z\ \int d\bbox x_2 \int
d\bbox x_1 \, \varphi_n^\dagger (\bbox x_2) \alpha_\mu 
{\cal G}(\bbox x_2,\bbox x_1, z)
 \alpha^\mu \varphi_n(\bbox x_1) \frac{e^{-bx_{21}}}{ x_{21}} -\delta m\int 
  d\bbox x\ \varphi_n^\dagger (\bbox x) \beta \varphi_n(\bbox x), 
\label{eq:highe}
\end{equation} 
\narrowtext \noindent where $b\equiv -i\left[({\cal E}_n-z)^2
  +i\epsilon\right]^{1/2}$, ${\rm Re}(b)>0$ is the photon energy, and
$x_{21} \equiv ||\bbox x_2 - \bbox x_1||$ ($\epsilon$ is an
infinitesimal positive quantity). In these expressions, $\varphi_n$
and ${\cal E}_n$ are the eigenfunction and eigenvalue of the Dirac
equation for the bound state $n$, and ${\cal G}$ is the Dirac
Green's function: ${\cal G}(z) = ({\cal H}-z)^{-1}$, where
${\cal H}=\bbox \alpha\cdot\bbox p + {\cal V} + \beta$ is the Dirac
Hamiltonian. Indices $l$ and $m$ are summed from 1 to 3, and index
$\mu$ is summed from 0 to 3.  The contour ${\rm C_H}$ extends from
$-i\infty$ to $0-i\epsilon$ and from $0+i\epsilon$ to $+i\infty$.

Since the high-energy part and the renormalization procedure described
in Refs.~\cite{moh82,mak92,iam92,iam98} are already known for
arbitrary angular momenta, this paper is concerned with the low-energy
part~(\ref{eq:lowe}). For a spherically symmetric potential ${\mathcal
  V}$, a separation of the photon propagator and of Dirac
wavefunctions into radial and angular parts yields the following
expression for the low-energy part of the self-energy:
\begin{eqnarray}
&&{\mathcal E}_L =\frac{\alpha}{\pi}{\mathcal E}_n-\frac{\alpha}{\pi}{\mathrm P}\int_0^{{\mathcal E}_n}dz 
\int_0^\infty dx_2 x_2^2 \int_0^\infty dx_1x_1^2 \nonumber\\
&&\times 
 \sum_\kappa\sum_{i,j=1}^2f_{\overline \imath}(x_2) G_\kappa^{ij} 
(x_2,x_1,z) f_{\overline \jmath}(x_1) A_{\kappa,\kappa_{n}}^{ij}(x_2,x_1),
\label{eq:loweint} \end{eqnarray} 
\noindent where the $f_i$'s are the radial components of
the wavefunction $\varphi_n$, $\bar{\imath} \equiv 3-i$,
$G_\kappa^{ij}$ are the radial components of the Green's functions,
and $A_{\kappa,\kappa_n}^{ij}$ are functions that contain angular
integrations as well as the photon propagator. The Dirac angular
quantum number of the electron for which the self-energy is calculated
is denoted by $\kappa_n$. Detailed definitions of these notations can
be found in Ref.~\cite{moh74a}.

Expression (\ref{eq:loweint}) can for instance be applied to a pure
Coulomb potential ${\mathcal V}$ or to the self-energy screening
correction~\cite{iam01}. The potential ${\mathcal V}$ affects the
radial components $f$ and $G$; on the contrary, the angular
coefficients $A$ are independent of ${\mathcal V}$; this paper
provides analytical formulas for these coefficients.

\section{Final formulas for the angular integrals}
\label{sec:form-angul}

In this section, we present our \emph{final formulas} for the
angular coefficients $A^{ij}_{\kappa,\kappa_n}$ introduced in
Eq.~(\ref{eq:loweint}). Our results can be directly implemented on a
computer.  Derivations of the results presented here can be found in
Sec.~\ref{sec:integA11} and Sec.~\ref{sec:integA12}.

We present successively our results for the coefficients
$A^{11}_{\kappa,\kappa_n}$ and $A^{12}_{\kappa,\kappa_n}$. The other two coefficients can be
obtained through the following symmetries~\cite{moh74a}:
\begin{eqnarray*}
A^{21}_{\kappa,\kappa_n} & = & A^{12}_{-\kappa,-\kappa_n}\\
A^{22}_{\kappa,\kappa_n} & = & A^{11}_{-\kappa,-\kappa_n}.
\end{eqnarray*}

\subsection{Final result for $A^{11}_{\kappa,\kappa_n}$}\label{sec:a11final}

\subsubsection{Initial expression and notations}

We give in this section the numerically optimized form that we found
for the angular coefficient $A^{11}_{\kappa,\kappa_n}$.  The most general formula
available for this coefficient is~\cite{moh74a}
\begin{eqnarray}
\nonumber
A^{11}_{\kappa,\kappa_n}
&\equiv&
|\kappa|
\int_{-1}^{1}\!
  d\xi \,
\bigg[
P_{l_{\kappa}}(\xi) P_{l_{\kappa_n-1}}(\xi) b^2 T(\rho)
\\
&&
{}-
\frac{1}{\kappa\kappa_n} (1-\xi^2) P'_{l_{\kappa}}(\xi)
P'_{l_{\kappa_n-1}}(\xi) \frac{1}{\rho}\frac{d}{d\rho} T(\rho)
\bigg],
\end{eqnarray}
where $P_n$ is the Legendre polynomial~\cite{aas72} of degree
$n$, and where
\begin{equation}\label{eq:defOrbital}
l_{\kappa} \equiv |\kappa+1/2|-1/2
\label{eq:a11general}
\end{equation}
denotes the \emph{orbital} angular momentum associated with the Dirac angular
quantum number $\kappa$; the other functions are defined as:
\begin{equation}\label{eq:defRho}
\rho(x_1,x_2,\xi)
\equiv\sqrt{x_1^2+x_2^2-2x_1x_2\xi}
\end{equation}
and
\begin{equation}\label{eq:defT}
T(\rho)\equiv
\sin(b\rho)/(b^2\rho).
\end{equation}
The quantities $\kappa_n$, $\kappa$, $x_1$, $x_2$ and $b$ are considered here
as fixed parameters. Since distances $x_1$ and $x_2$ often appear
multiplied by the energy $b$, we define
\[
y_1 \equiv b\, x_1 \mbox{ and } y_2 \equiv b\, x_2.
\]

It is also convenient to introduce the notation 
\[
{\mathcal J}^2_{l} \equiv l(l + 1)
\]
for the eigenvalues of the squared angular momentum operator in terms
of the angular momentum $l$.

\subsubsection{Result of the analytical integration}

With the above notations, our result for the integration
in~Eq.~(\ref{eq:a11general}) reads:
\begin{eqnarray}
\nonumber
A^{11}_{\kappa,\kappa_n}
&=&
\sum_{\makebox[0pt]{$\scriptstyle
    l=|l_{\kappa}-l_{-\kappa_n}|$}}^{\makebox[0pt]{$\scriptstyle l_{\kappa}+l_{-\kappa_n}$}}
(2l+1)|\kappa|
\Bigg[
 2 b
\left(
\begin{array}{ccc}
l_{\kappa} & l_{-\kappa_n} & l\\ 
0 & 0 & 0
\end{array}
\right)
^2
 j_{l}(y_1) j_{l}(y_2)
\\
\label{eq:a11final}
&&
\qquad
+
\frac{1}{ \kappa\kappa_n}
{\mathcal L}(l,l_{\kappa},l_{-\kappa_n})
 \frac{j_{l}(y_1)}{y_1}
 \frac{j_{l}(y_2)}{x_2}
\Bigg],
\end{eqnarray}
where the $2\times 3$ matrices are Wigner $3j$-coefficients, where the
$j_{l}$'s are the spherical Bessel functions of the first
kind~\cite{aas72}, and where we have introduced a symbol for
the following quantity:
\widetext
\begin{equation}\label{eq:resultLegendre}
{\mathcal L}(a,b,c)
\equiv
\left\{
\begin{array}{l}
\displaystyle
\sum_{
\scriptsize
\parbox{20ex}{
\centering

$i=|a-b|$\\
$i+|a-b|$ even

}}^{c-1}
\left(
\begin{array}{ccc}
a & b & i\\ 
0 & 0 & 0
\end{array}
\right)
^2(2i+1)
 ({\mathcal J}^2_{a} + {\mathcal J}^2_{b} - {\mathcal J}^2_{i})
\quad
\mbox{if $a+b+c$ odd}
\\
\\
0
\quad
\mbox{otherwise.}
\end{array}
\right.
\end{equation}
\narrowtext\noindent
Despite the non-symmetrical form (\ref{eq:resultLegendre}), the
quantity ${\mathcal L}(a,b,c)$ is totally \emph{symmetrical}
with respect to any permutation of its arguments, as we prove in
Sec.~\ref{sec:integDerivLeg}.

\subsubsection{Numerical implementation}

Equation~(\ref{eq:a11final}) can be readily implemented on a computer.
In Eq.~(\ref{eq:a11final}), no term is \emph{singular} (with respect to the
parameters $b$, $x_1$ and $x_2$). In fact, the expansion of the Bessel
functions about the origin~\cite{aas72} is such that
$j_{l}(z) = {\mathcal O}(z^l)$; singularities could appear only
through $j_{0}(z)/z$, which is singular as $z\rightarrow 0$.  But
$l=0$ imposes that $l+l_{\kappa}+l_{-\kappa_n}$ be even, so that
$j_{0}(z)/z$ is actually never used in Eq.~(\ref{eq:a11final}).

Since the number of terms in Eq.~(\ref{eq:resultLegendre}) is smaller than $c$,
we have decided to set $c=l_{-\kappa_n}$ in Eq.~(\ref{eq:a11final}), because
$l_{-\kappa_n}$ is small, being the orbital momentum of the level for
which the self-energy is calculated~\cite{moh74a}. Furthermore, we
note that in~Eq.~(\ref{eq:a11final}), for a given $l$, only \emph{one} term
inside the brackets contribute to $A^{11}_{\kappa,\kappa_n}$ because of 
\emph{parity} selection on Wigner $3j$-coefficients
\cite{edm96} and in our quantity ${\mathcal L}(a,b,c)$
of~Eq.~(\ref{eq:resultLegendre}).

We have also checked whether result (\ref{eq:a11final}) could
  yield strong \emph{numerical cancellations} in the following
  ranges of parameters
\begin{equation} \label{eq:zone}
\begin{array}{c}
b\in [0,1]
,\,
x_1, x_2 \in [0,100]
,
\\
|\kappa_n| \in [1,10]
\mbox{\ and\ } |\kappa| \in [1,1000],
\end{array}
\end{equation}
which are typical of the values used in numerical
calculations~\cite{moh74a}. We checked for possible cancellations with
an arbitrary precision in the following two cases: (a)~two numbers add
up to a very small one (about $10^{-8}$ smaller than the two numbers);
(b)~a large number is added to a number which is much smaller (by a
factor of $10^{-9}$). Our checks have shown that \emph{no} such
numerical problems arise with the terms of Eq.~(\ref{eq:a11final}) when the
summation is done by starting from the higher limit
$l_{\kappa}+l_{-\kappa_n}$; this choice of summation order is
numerically motivated by the over-exponential damping of the spherical
Bessel
functions of the first kind~\cite{mps98}:
\begin{equation}\label{eq:besselDamping}
j_{l}(z) \sim  \frac{1}{2^{3/2} l }
{\left(\frac{ez}{2l}\right)}^l
\end{equation}
when $l > ez/2$; thus, summing the terms of the final
  result~(\ref{eq:a11final}) with \emph{decreasing} $l$'s allows
  one to include larger and larger terms in the total sum, which is
  necessary for obtaining accurate numerical sums.

We have numerically tested formula (\ref{eq:a11final}) against
\emph{Mathematica}, against the alternate formula presented below
[Eq.~(\ref{eq:a11kp})] and against Mohr's implementation of his special cases
$|\kappa_n|\le 2$; we have found excellent accuracy in the parameter
ranges used in the low-energy part calculations of the
self-energy~\cite{moh74a,moh82,mak92}, \emph{i.e.}, in the range of
parameters reached by our calculations of the self-energy in ions with
$Z\ge 10$:
\begin{equation}\label{eq:largeZone}
\begin{array}{c}
b\in [0,1]
,\,
x_1, x_2 \in [0,5000]
,
\\
|\kappa_n| \in [1,5]
\mbox{\  and\ } |\kappa| \in [1,5000].
\end{array}
\end{equation}

\subsection{Final result for $A^{12}_{\kappa,\kappa_n}$}

\subsubsection{General integration result}

We give in this section a form of the angular coefficient $A^{12}_{\kappa,\kappa_n}$
which is optimized for numerical calculations. We start from the
\emph{most general} formula available~\cite{moh74a} for this
angular coefficient:
\widetext
\begin{eqnarray}
\nonumber
A^{12}_{\kappa,\kappa_n} &\equiv&
-|\kappa|
\int_{-1}^{1}\!
  d\xi \,
\bigg\{\left[
P_{l_{\kappa-1}}(\xi) P_{l_{\kappa_n}}(\xi)
-\frac{1}{\kappa\kappa_n}(1-\xi^2)P'_{l_{\kappa-1}}(\xi) P'_{l_{\kappa_n}}(\xi)\right]
\frac{1}{\rho}\frac{d}{d\rho} T(\rho)\\
\label{eq:a12general}
&&\qquad\qquad\quad
+[x_2 P_{l_{\kappa-1}}(\xi)-x_1 P_{l_{\kappa}}(\xi)]
 [x_2 P_{l_{\kappa_n}}(\xi)-x_1 P_{l_{\kappa_n-1}}(\xi)]
\frac{1}{\rho}\frac{d}{d\rho}\frac{1}{\rho}\frac{d}{d\rho} T(\rho)
\bigg\}.
\end{eqnarray}
With the help of the notations defined previously, the (almost) final
result of our calculation of Eq.~(\ref{eq:a12general}) reads
\begin{eqnarray}
\nonumber
A^{12}_{\kappa,\kappa_n} &=&
\sum_{
\scriptsize
\parbox{20ex}{
\centering

$l=|l_{\kappa}-l_{\kappa_n}|$\\
  $l+l_{\kappa}+l_{\kappa_n}$ even
}}
^{l_{\kappa}+l_{\kappa_n}}
\frac{j_{l}(y_1)}{y_1} \frac{j_{l}(y_2)}{x_2}
\frac{|\kappa|}{\kappa \kappa_n} (2l+1)
\Bigg\{
\left(
\begin{array}{ccc}
l_{\kappa_n} & l_{\kappa} & l\\ 
0 & 0 & 0
\end{array}
\right)
^2
[{\mathcal J}^2_{\kappa_n} ({\mathcal J}^2_{l} + {\mathcal J}^2_{\kappa} - {\mathcal J}^2_{\kappa_n}) + \kappa^2 ({\mathcal J}^2_{l} -
{\mathcal J}^2_{\kappa} + {\mathcal J}^2_{\kappa_n})
]
\\
\nonumber
&&
\qquad \quad
{}
-
{\mathcal L}(l,l_{-\kappa},l_{\kappa_n})
\Bigg\}
\\
&&
\nonumber
\qquad
{}+ |\kappa| (2l+1)
\left(
\begin{array}{ccc}
l_{\kappa_n} & l_{\kappa} & l\\ 
0 & 0 & 0
\end{array}
\right)
^2
\times
\Bigg[
j_{l}'(y_1)\frac{j_{l}(y_2)}{x_2}
({\mathcal J}^2_{l}-{\mathcal J}^2_{\kappa}+{\mathcal J}^2_{\kappa_n})/\kappa_n
\nonumber
\\
&&
\qquad \quad {}+
\frac{j_{l}(y_1)}{x_1} j_{l}'(y_2)
({\mathcal J}^2_{l}+{\mathcal J}^2_{\kappa}-{\mathcal J}^2_{\kappa_n})/\kappa
\label{eq:a12final}
+
j_{l}'(y_1) j_{l}'(y_2)\times 2b
\Bigg],
\end{eqnarray}
\narrowtext
\noindent which is the generalization to \emph{any} atomic state of the
angular coefficient $A^{12}_{\kappa,\kappa_n}$ of the low-energy part of the
self-energy~\cite{moh74a}. Equation~(\ref{eq:a12final}) has a few
crucial computational advantages: (a)~all the sums contain a
\emph{finite} number of terms; (b)~there are no large terms due
to small $b$, $x_1$ or $x_2$; in fact, it is easy to see that the only
divergent quantities can come from $j_{l=0}(z)/z$, but that this
quantity is actually always multiplied by 0. We thus gain
  numerical accuracy compared to the published
  formulas~\cite{moh82,mak92} for the special cases $\kappa_n=\pm 2$.

\subsubsection{Numerical optimization}

As for $A^{11}_{\kappa,\kappa_n}$, we have checked whether result
(\ref{eq:a12final}) yields strong \emph{numerical cancellations}
in the parameter ranges defined by (\ref{eq:zone}). Our checks have
shown that \emph{no} such numerical problems arise with the terms
of Eq.~(\ref{eq:a12final}) [evaluated in the order indicated by the parentheses,
with summations done from the higher limit because of
Eq.~(\ref{eq:besselDamping})], \emph{except} sometimes for the minimum $l$,
\emph{i.e.}, for
\[
l=l_{\mathrm min}\equiv|l_{\kappa} - l_{\kappa_n}|.
\]

However, we have cured this disease by noticing patterns in the way
numerical cancellation appear: we have experimentally found four
different areas of the $(\kappa,\kappa_n)$ plane, defined by which of the terms
of Eq.~(\ref{eq:a12final}) cancel each other; the ``terms'' are defined here by
separating terms with zero, one and two derivatives of Bessel
functions: we select the four terms of Eq.~(\ref{eq:a12final}) associated with
$[j_{l_{\mathrm min}}(y_1)/y_1 ][j_{l_{\mathrm min}}(y_2)/x_2]$,
$j_{l_{\mathrm min}}'(y_1) [j_{l_{\mathrm min}}(y_2)/x_2]$,
$[j_{l_{\mathrm min}}(y_1)/x_1] j_{l_{\mathrm min}}'(y_2)$, and
$j_{l_{\mathrm min}}'(y_1) j_{l_{\mathrm min}}'(y_2)$.

By using explicit polynomial formulas for the orbital angular momenta
$l_{\kappa}$ and $l_{\kappa_n}$ in each region, it is possible to
obtain simplified expressions for the four terms of Eq.~(\ref{eq:a12final}); 
terms can then generally be grouped together [in a way that depends on
the area in which $(\kappa,\kappa_n)$ lies] through the 
 Bessel identity~\cite{aas72}
\begin{equation}\label{eq:besselCancel}
\frac{l}{z} j_{l}(z) - j_{l}'(z) = j_{l+1}(z).
\end{equation}
We have found numerically that this identity can 
yield  very
strong \emph{numerical cancellations} between 
the $l=l_{\mathrm min}$ terms of Eq.~(\ref{eq:a12final}).

Thus, we have mathematically found the equations of the four
cancellation areas, and we have used identity~(\ref{eq:besselCancel})
in order to express the final result in terms of the \emph{right}-hand
side of Eq.~(\ref{eq:besselCancel}) instead of numerically calculating the
\emph{left}-hand side of the identity.  The precise shapes of the
areas in the $(\kappa,\kappa_n)$ plane are
\begin{equation}
\begin{array}{ll}
\mbox{area 1:}
& \kappa\kappa_n < 0 \mbox{ and } \kappa(l_{\kappa}-l_{\kappa_n}) \le
0
\\
\mbox{area 2:}
&
\kappa\kappa_n < 0 \mbox{ and } \kappa(l_{\kappa}-l_{\kappa_n}) > 0
\\
\mbox{area 3:}
&
\kappa\kappa_n > 0 \mbox{ and } \kappa(l_{\kappa}-l_{\kappa_n}) \le 0
\\
\mbox{area 4:}
&
\kappa\kappa_n > 0 \mbox{ and } \kappa(l_{\kappa}-l_{\kappa_n}) > 0.
\end{array}
\end{equation}
A numerical implementation of Eq.~(\ref{eq:a12final}) should thus calculate
the terms corresponding to $l=l_{\mathrm min}$ with the following expressions,
which depend on the area in which $(\kappa,\kappa_n)$ lies:
\widetext
\begin{equation}
A^{12}_{\kappa,\kappa_n}(l = l_{\mathrm min})
=
\left\{
\begin{array}{ll}
2b|\kappa| (2l_{\mathrm min} + 1)
\left(
\begin{array}{ccc}
l_{\kappa_n} & l_{\kappa} & l_{\mathrm min}\\ 
0 & 0 & 0
\end{array}
\right)
^2
j_{l_{\mathrm min}+1}(y_1) j_{l_{\mathrm min}+1}(y_2)
&
\mbox{in area 1}
\\
A^{12}_{\kappa,\kappa_n}(l=l_{\mathrm min}) \mbox{ from Eq.~(\ref{eq:a12final})}
&
\mbox{in area 2}
\\
\left(
\begin{array}{l}
- |\kappa| (2l_{\mathrm min} + 1)
\left(
\begin{array}{ccc}
l_{\kappa_n} & l_{\kappa} & l_{\mathrm min}\\ 
0 & 0 & 0
\end{array}
\right)
^2
j_{l_{\mathrm min}+1}(y_1)
\\
\qquad\qquad
\times
\left\{
[ j_{l_{\mathrm min}}(y_2)/x_2] ({\mathcal J}^2_{l_{\mathrm min}} - {\mathcal J}^2_{\kappa} + {\mathcal J}^2_{\kappa_n})/\kappa_n
+
2b j_{l_{\mathrm min}}'(y_2) 
\right\}
\end{array}
\right)
&
\mbox{in area 3}
\\
\left(
\begin{array}{l}
- |\kappa| (2l_{\mathrm min} + 1)
\left(
\begin{array}{ccc}
l_{\kappa_n} & l_{\kappa} & l_{\mathrm min}\\ 
0 & 0 & 0
\end{array}
\right)
^2
j_{l_{\mathrm min}+1}(y_2)
\\
\label{eq:lMinTerm}
\qquad\qquad
\times
\left\{
 [j_{l_{\mathrm min}}(y_1)/x_1] ({\mathcal J}^2_{l_{\mathrm min}} + {\mathcal J}^2_{\kappa} - {\mathcal J}^2_{\kappa_n})/\kappa
+
2b j_{l_{\mathrm min}}'(y_1) 
\right\}
\end{array}
\right)
&
\mbox{in area 4}.
\end{array}
\right.
\end{equation}

\narrowtext

Furthermore, no factor of the form $j_{0}(z)/z$, which
\emph{diverges} as $z\rightarrow 0$ appears anymore in our formulas for
$A^{12}_{\kappa,\kappa_n}$: such a factor can only be found in the $l=l_{\mathrm min}$ term
of result (\ref{eq:a12final}); but Eq.~(\ref{eq:lMinTerm}) must be used for this
term. $l_{\mathrm min}\equiv |l_{\kappa}-l_{\kappa_n}|$ is obviously
never encountered in the first two areas of Eq.~(\ref{eq:lMinTerm}). In the last
two areas of Eq.~(\ref{eq:lMinTerm}), $j_{0}(z)/z$ is used but multiplied
by a factor \emph{zero}, as can be seen by using the fact that
$\kappa\kappa_n>0$.

In summary, Eqs.~(\ref{eq:a12final}) and (\ref{eq:lMinTerm}) give numerically
optimized formulas for the angle coefficient $A^{12}_{\kappa,\kappa_n}$ of
Eq.~(\ref{eq:a12general}). We have numerically tested these formulas against
\emph{Mathematica}, against the alternate formula presented below
[Eq.~(\ref{eq:a12paulFinal})], and against Mohr's implementation of his special
cases $|\kappa_n|\le 2$; we have found excellent accuracy in the
physical parameter ranges~(\ref{eq:largeZone}).

\section{Numerical results for the self-energy}

\label{sec:numer-res}

The formulas (\ref{eq:a11final}), (\ref{eq:a12final}) and
(\ref{eq:lMinTerm}) presented above allow us to numerically evaluate
the QED self-energy contribution ${\mathcal E}_{\mathrm L}$ formally
expressed in Eq.~(\ref{eq:loweint}). The other self-energy contribution
[${\mathcal E}_{\mathrm H}$, Eq.~(\ref{eq:highe})] can be calculated for an
arbitrary electronic state by means of previously published
methods~\cite{moh82}. We present in this section numerical evaluations
of the self-energy of atomic electrons with principal quantum number
$3\le n\le 5$ and with angular number $3\le |\kappa| \le 5$, in
hydrogenlike ions with nuclear charge $60\le Z\le 110$. We used a
Fortran implementation of formulas (\ref{eq:a11final}),
(\ref{eq:a12final}) and (\ref{eq:lMinTerm}), as well as previously
published numerical
procedures~\cite{moh74b,moh82,mak92,iam92,iam98,iam01,iam95,iam91}.

\begin{figure}
\psfig{figure=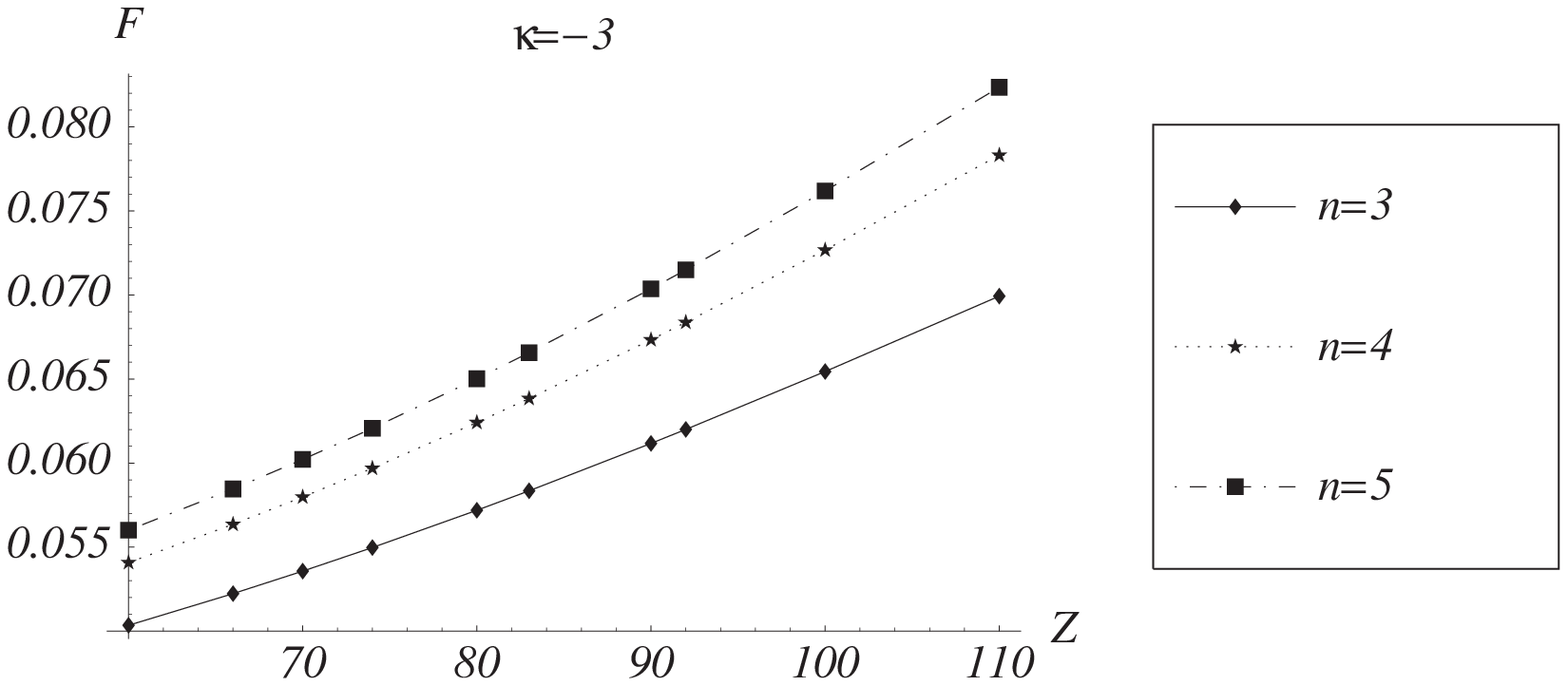}
\caption{\label{fig:firstKfixed}Scaled self-energy $F(Z\alpha)$ for $d_{5/2}$ states of levels $n=3$, 4
  and 5.}
\end{figure}

\begin{figure}
\psfig{figure=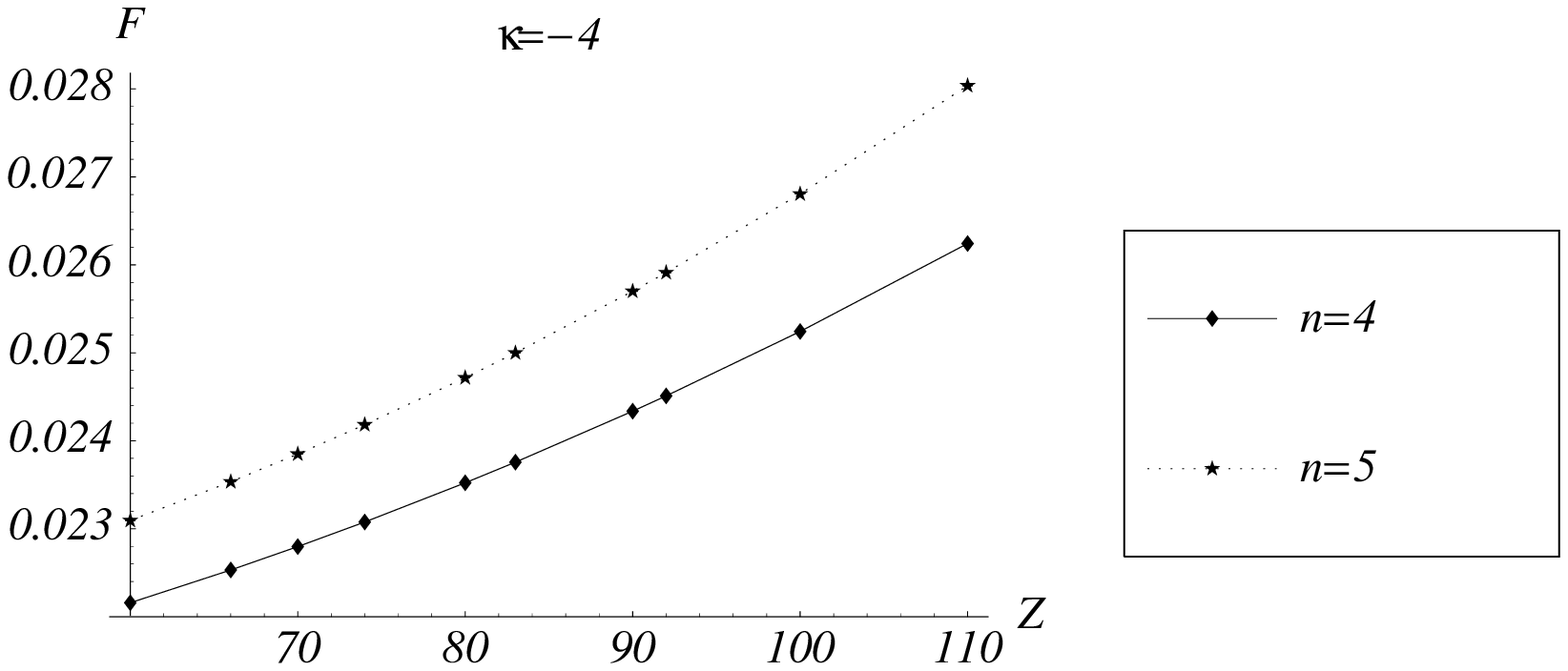}
\caption{\label{fig:lastKfixed}Scaled self-energy $F(Z\alpha)$ for $f_{7/2}$ states of levels $n$=4
  and 5.}
\end{figure}

\begin{figure}
\psfig{figure=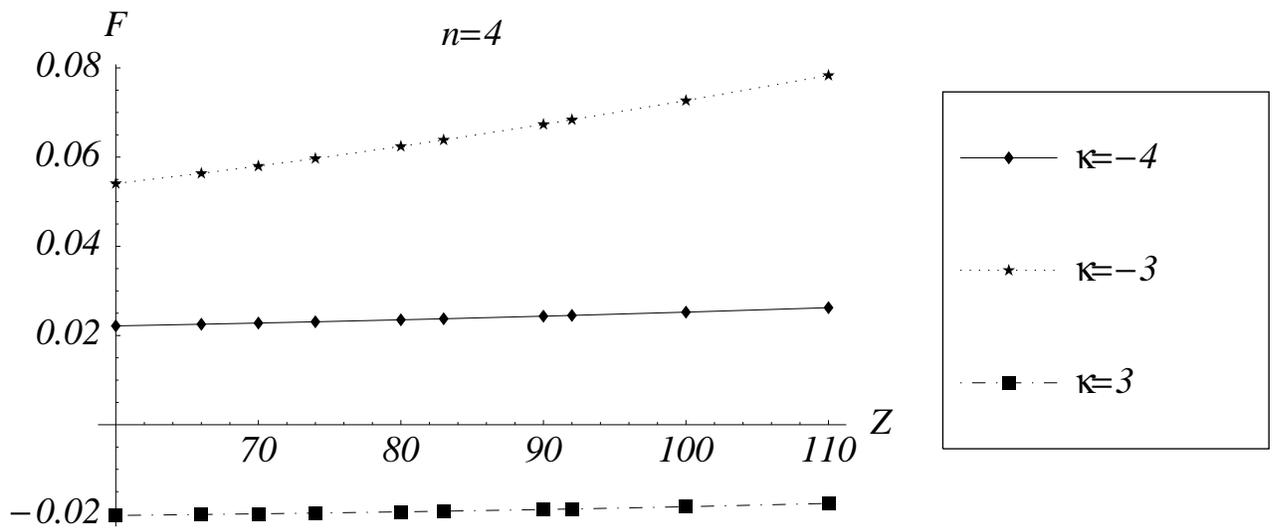}
\caption{\label{fig:firstNfixed}Scaled self-energy $F(Z\alpha)$ for atomic level $n=4$ and angular
  quantum numbers $\kappa=-3$, 3 and -4.
}
\end{figure}

\begin{figure}
\psfig{figure=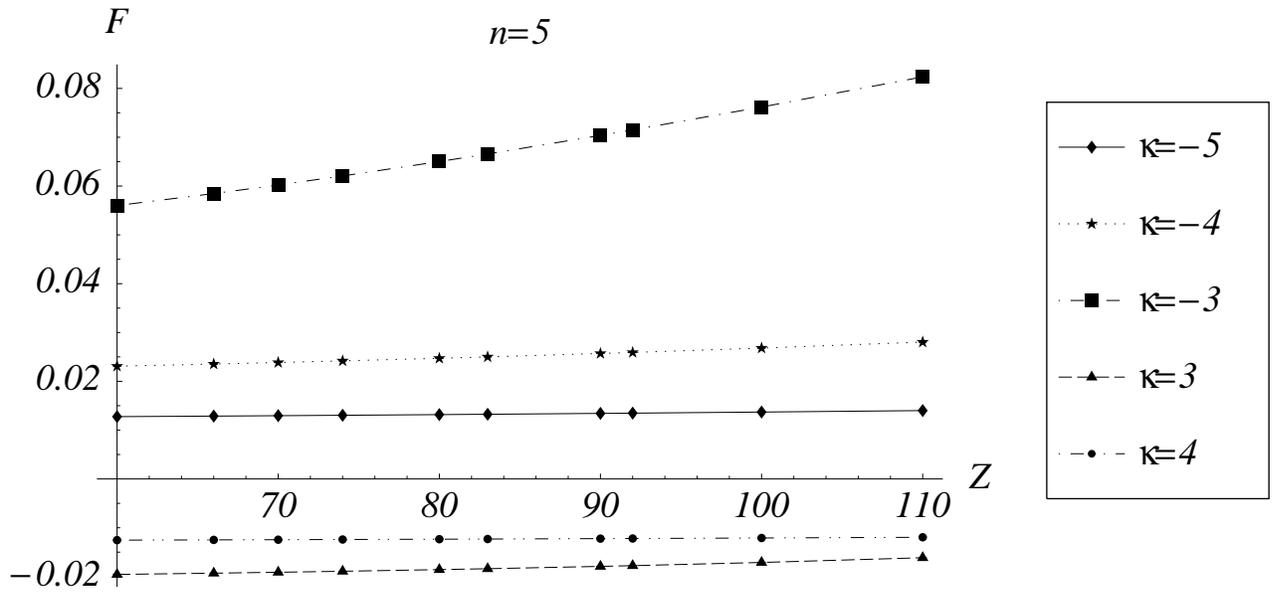}
\caption{\label{fig:lastNfixed}Scaled self-energy $F(Z\alpha)$ for atomic level $n=5$ and angular
  quantum numbers $\kappa$ such that $|\kappa| \ge 3$.}
\end{figure}

Numerical results are most conveniently expressed in terms of the usual
\emph{scaled} self-energy $F$:
\begin{equation}
\label{eq:scaledSE}
{\mathcal E}_{\mathrm SE}
\equiv
\frac{\alpha}{\pi}
\frac{(Z\alpha)^4}{n^3}
F(Z\alpha){m_{\mathrm e} c^2},
\end{equation}
where ${\mathcal E}_{\mathrm SE}$ is the self-energy shift
(\ref{eq:defSE}). Our evaluations of the scaled value $F$ are
presented in Tables~\ref{tbl:firstTable}--\ref{tbl:lastTable}.
When represented as a function of $Z\alpha$, our values display smooth
curves, as can be seen in Figs.~\ref{fig:firstKfixed}--\ref{fig:lastNfixed}.

From Figs.~\ref{fig:firstKfixed}--\ref{fig:lastKfixed}, we notice that
the \emph{scaled} self-energy does not greatly depend on the atomic
level $n$; put in other words, the $n$-dependence of the self-energy
is quite well \emph{captured by the scaling $1/n^3$} of Eq.~(\ref{eq:scaledSE}),
as has been observed for $s$ and $d_{3/2}$ levels~\cite{iam98a}.
On the practical side, the slow variation of the scaled self-energy
$F$ with respect to $n$ allows one to use the values we present here
as estimates for higher levels $n$.

We have also grouped self-energy values by atomic level $n$ in
Figs.~\ref{fig:firstNfixed} and \ref{fig:lastNfixed}.  We notice that
states with \emph{identical orbital angular momenta} have self-energy
curves that look \emph{parallel} on the scale we used.  This property
can be understood for \emph{low} nuclear charge $Z$, since states
other than $s$ states have the property that $F$ is dominated in this
region by a function of the form
\[
A_{40} + (Z\alpha)^2 \log[(Z\alpha)^{-2}]  A_{61} 
\]
for $Z\alpha \ll 1$, and where $A_{40}$ and $A_{61}$ are the usual
coefficients of the semi-analytic expansion of the self-energy
shift~\cite{pac98}. (The $(Z\alpha)^2 A_{60}$ contribution is 
negligible for $Z\alpha \rightarrow 0$: $(Z\alpha)^2 =
o[(Z\alpha)^2\log([Z\alpha]^{-2})]
$.)  It turns out that for
$d,f,g$\ldots\  states, $A_{61}$ depends only on the \emph{orbital}
angular momentum (This is due to the smooth behavior of the
wavefunction of such states at the origin~\cite{jen}.); as a
consequence, the self-energy curves for states with identical quantum
numbers $n$ and $l$ (angular momentum) are parallel in the limit
$Z\alpha \rightarrow 0$. For the higher $Z$ values of the results we
present here, the expansion of $F$ in $Z\alpha$ is not supposed to
hold; however our numerical values show that the difference between
the scaled self-energies of two levels with the same $n$ and $l$
varies only on the level of a few percents in the range $60 \le Z \le
110$. Precisely, the difference between the self-energy $F$ for two
states with identical $n$ and $l$ is well approximated by the value of
this difference at the limit $Z\alpha \rightarrow 0$, which is known
to be~\cite{das90}
\begin{equation}\label{eq:splittingSameL}
\lim_{Z\alpha \rightarrow 0}
[
F_{n,\kappa=-(l+1)}(Z\alpha)
-
F_{n,\kappa=l}(Z\alpha)
]
=
\frac{1}{l(l+1)},
\end{equation}
which approximates (to the level of a few percents) the splitting
between states with identical $n$ and $l$ present in our results; it
would be interesting to find an explanation of this property of the
scaled self-energy $F$ at high $Z$. At low $Z$, the splitting
(\ref{eq:splittingSameL}) is due to the anomalous magnetic moment of
the electron~\cite{das90}, and maybe does this effect dominate the
splitting we observe between states of identical $n$ and $l$ in our
high-$Z$ results.

Our results are coherent with the numerical results published by
Yerokhin and Shabaev~\cite{v99} for nuclear charges $Z=74$, $83$ and
$92$, as shown in Tables~\ref{tbl:compYerokhinFirst} and
\ref{tbl:compYerokhinLast}. However, we note that our evaluations of
the self-energy often lie \emph{below} the values of~\cite{v99}, and
that our results have uncertainties smaller by about two orders of
magnitude. Furthermore, the value that lies the furthest from their
error bars is that of the $5d_{5/2}$ level for $Z=74$, which is
located relatively close ($-1.2$ standard deviations) from the
previously published result~\cite{v99}.

\section{Calculation of the angular coefficients: method
for optimized numerical evaluation}
\label{sec:optnumder}

  Our analytical evaluations of $A^{11}_{\kappa,\kappa_n}$ and
  $A^{12}_{\kappa,\kappa_n}$~\cite{moh74a} rest on a few cornerstones.
  
  First, the dependence on $\xi$ of the integrands of Eqs.~(\ref{eq:a11general})
  and (\ref{eq:a12general}) is made simple by expanding the $T$
  function with partial waves. In fact, in Eq.~(\ref{eq:defRho}), $\rho$ is
  the distance between the two interaction points of the self-energy,
  $\xi$ being the cosine of the angle between these two
  points~\cite{moh74a}; the $T$ function of Eq.~(\ref{eq:defT}) can
  therefore be expanded in partial
  waves~\cite{aas72}:
\begin{equation}\label{eq:devT}
  T[\rho(x_1,x_2,\xi)] = \frac{1}{b} \sum_{l \ge 0} (2l+1)
P_{l}(\xi) j_l(y_1) j_l(y_2),
\end{equation}
where we still have
\[
y_1\equiv b x_1 \mbox{ and } y_2\equiv b x_2.
\]
Thus, integrals of terms of Eqs.~(\ref{eq:a11general}) and (\ref{eq:a12general})
which do not contain derivatives of $T$ are simply integrations of
\emph{polynomials} in $\xi$. And since $d /
d \rho$ can be calculated with $\partial / \partial \xi$
[see Eq.~(\ref{eq:differentiationRho}) below], all terms are as well integrals
of polynomials.

Second, integrations of polynomials can be expanded in a simple form
with the help of expansions on an orthogonal basis. Much of our work
as thus been concentrated on obtaining formulas with polynomials in
$\xi$ that can be simply expanded onto the basis of Legendre polynomials.

Third, the sum over partial waves in Eq.~(\ref{eq:devT}) contains an
\emph{infinite} number of terms. Numerical calculations of
angular coefficients $A^{11}_{\kappa,\kappa_n}$ and $A^{12}_{\kappa,\kappa_n}$ are more
accurate if only a finite number of these terms contribute to the
final result. This can be achieved by obtaining expressions in which
we can apply triangular identities on Legendre polynomials (Legendre
polynomials are a special kind of spherical harmonics); such a goal
has driven our search of numerically efficient expressions for
$A^{11}_{\kappa,\kappa_n}$ and $A^{12}_{\kappa,\kappa_n}$.

Accordingly to these principles, expansions and scalar products in
Eqs.~(\ref{eq:fundamentalIntegration}), (\ref{eq:beforeScalarProduct}),
(\ref{eq:fundamentalIntegration2:final}), and (\ref{eq:scalarProdLeg})
are especially important to the following calculations.

\subsection{Integration in $A^{11}_{\kappa,\kappa_n}$}
\label{sec:integA11}

\subsubsection{First steps}

We describe in this section our calculation~(\ref{eq:a11final}) of the
integral over $\xi$ in the original expression of the angular
coefficient~$A^{11}_{\kappa,\kappa_n}$ of Eq.~(\ref{eq:a11general}).

Following Mohr~\cite{moh73}, we evaluate the derivative $1/\rho
\times d/d\rho$ of $A^{11}_{\kappa,\kappa_n}$
in~Eq.~(\ref{eq:a11general}) by using a differentiation with respect to the
$\xi$ parameter of the $\rho$ function of Eq.~(\ref{eq:defRho}): in fact, for
any function $f$, the definition of $\rho$ implies that
\begin{equation}\label{eq:differentiationRho}
\left.\frac{1}{\rho}\frac{d}{d\rho} f(\rho)\right|_{\rho(x_1,x_2,\xi)}
= 
\frac{-1}{x_1 x_2}
\frac{\partial}{\partial \xi}
f[\rho(x_1,x_2,\xi)].
\end{equation}

Starting from Eq.~(\ref{eq:a11general}), a first step consists in using the
\emph{partial-wave expansion}~(\ref{eq:devT}) of $T$ function
defined in~Eq.~(\ref{eq:defT}). With Eqs.~(\ref{eq:devT}) and
(\ref{eq:differentiationRho}), we thus obtain a form of $A^{11}_{\kappa,\kappa_n}$ in
which products of three Legendre polynomials (or derivatives of
Legendre polynomials) must be integrated:
\widetext
\begin{equation}\label{eq:a11:1}
A^{11}_{\kappa,\kappa_n} =
|\kappa|
\sum_{l \ge 0}
(2l+1) j_l(y_1) j_l(y_2)
\int_{-1}^{1}\!
  d\xi \,
\left[
b
P_{l_{\kappa}}(\xi) P_{l_{-\kappa_n}}(\xi)
P_{l}(\xi)
-
\frac{1}{\kappa\kappa_n}
\frac{-1}{x_1 x_2}
 (1-\xi^2) P'_{l_{\kappa}}(\xi)
P'_{l_{-\kappa_n}}(\xi)
P'_{l}(\xi)
\right],
\end{equation}
\narrowtext\noindent
where we have used the identity of orbital
momenta $l_{\kappa_n- 1} = l_{-\kappa_n}$.

We can integrate over $\xi$ in the first term of expression
(\ref{eq:a11:1}) with  the well-known result~\cite{edm96}
\begin{equation}\label{eq:fundamentalIntegration}
\int_{-1}^{1}\!
  d\xi \,
P_{a}(\xi)P_{b}(\xi)P_{c}(\xi)
=2
\left(
\begin{array}{ccc}
a & b & c\\ 
0 & 0 & 0
\end{array}
\right)
^2.
\end{equation}

\subsubsection{Integral of derivatives of Legendre polynomials}
\label{sec:integDerivLeg}

Integration of the second term of $A^{11}_{\kappa,\kappa_n}$ in Eq.~(\ref{eq:a11:1}) is more
difficult. We have found that
\begin{equation}\label{eq:fundamentalIntegration2}
\int_{-1}^{1}\!
  d\xi \,
(1-\xi^2)
P'_{a}(\xi)P'_{b}(\xi)P'_{c}(\xi)
=
{\mathcal L}(a,b,c),
\end{equation}
where ${\mathcal L}(a,b,c)$ is defined in Eq.~(\ref{eq:resultLegendre}). The
idea behind the derivation of this result was to consider the product
$(1-\xi^2) P'_{a}(\xi)P'_{b}(\xi)$ as two coupled angular
momentum eigenstates, so as to decompose this quantity onto the basis
of Legendre polynomials, which we do now.  The derivatives of Legendre
polynomials are related to Legendre polynomials of orders~$-1$ and~$1$
by the following identities~\cite{edm96}:
\begin{equation}\label{eq:legendreDer:legendreOrder}
\sqrt{1-\xi^2}P'_{i}(\xi) = P_{i}^{1}(\xi)
= - {\mathcal J}^2_{i} P_{i}^{-1}(\xi)
,
\end{equation}
and the Legendre polynomials of a given order and degree are directly
related to spherical harmonics~\cite{edm96}:
\begin{equation}\label{eq:sphericalHarm:legendreOrder}
Y_{l, m}(\theta, \phi)
=
{(-1)}^m \sqrt{\frac{(2l+1) (l-m)!}{4\pi (l+m)!}}
P_{l}^{m}(\cos \theta) e^{i m \phi}
,
\end{equation}
where the spherical harmonics $Y_{l, m}(\theta, \phi)$ is defined with
the convention of Edmonds~\cite{edm96}, and where
$P_{l}^{m}(\xi)$ is the Legendre polynomial of degree $l$ and
order $m$~\cite{aas72}. We thus see that the left-hand side of
Eq.~(\ref{eq:fundamentalIntegration2}) can be rewritten with spherical
harmonics: in fact, we have from Eqs.~(\ref{eq:legendreDer:legendreOrder})
and~(\ref{eq:sphericalHarm:legendreOrder}) that
\begin{eqnarray}
\lefteqn{(1-\cos^2 \theta) P'_{a}(\cos
  \theta)P'_{b}(\cos \theta)}
\quad&
\nonumber
\\
&=&
-{\mathcal J}^2_{b} P_{a}^{1}(\cos \theta)
P_{b}^{-1}(\cos \theta)
\nonumber
\\
&=&
4\pi \sqrt{\frac{{\mathcal J}^2_{a}{\mathcal J}^2_{b}}{(2a+1)(2b+1)}}
Y_{a, 1}(\theta, \phi) Y_{b, -1}(\theta, \phi);
\label{eq:toSphericalHarm}
\end{eqnarray}
the right-hand side of the above expression is constant with respect
to $\phi$ because the product of spherical harmonics has an orbital
momentum projection of $0$. The two spherical harmonics of
Eq.~(\ref{eq:toSphericalHarm}) can be coupled~\cite{edm96}:
\widetext
\begin{equation}\label{eq:coupledSphericalHarm}
Y_{a, 1}(\theta, \phi) Y_{b, -1}(\theta, \phi)
=
\sum_{i \ge 0} \sqrt{\frac{(2a+1)(2b+1)(2i+1)}{4\pi}}
\left(
\begin{array}{ccc}
a & b & i\\ 
1 & -1 & 0
\end{array}
\right)
\left(
\begin{array}{ccc}
a & b & i\\ 
0 & 0 & 0
\end{array}
\right)
Y_{i, 0}(\theta, \phi).
\end{equation}

Thus, the quantity $(1-\xi^2)P'_{a}(\xi)P'_{b}(\xi)$ of
Eq.~(\ref{eq:fundamentalIntegration2}) can be expanded over Legendre
polynomials: with Eqs.~(\ref{eq:toSphericalHarm}),
(\ref{eq:coupledSphericalHarm}) and
(\ref{eq:sphericalHarm:legendreOrder}), we obtain that
\begin{equation}\label{eq:beforeScalarProduct}
(1-\xi^2)
P'_{a}(\xi)P'_{b}(\xi)
=
-
\sqrt{{\mathcal J}^2_{a}{\mathcal J}^2_{b}} \sum_{i\ge 0}
(2i+1)
\left(
\begin{array}{ccc}
a & b & i\\ 
1 & -1 & 0
\end{array}
\right)
\left(
\begin{array}{ccc}
a & b & i\\ 
0 & 0 & 0
\end{array}
\right)
P_{i}(\xi).
\end{equation}
\narrowtext\noindent
At this point, it is useful to have a closer look at parity selection
rules in Eq.~(\ref{eq:beforeScalarProduct}).  Since only terms with $a+b+i$ even
contribute to the sum over $i$ in Eq.~(\ref{eq:beforeScalarProduct}) (because of
the second $3j$-coefficient), we can use the following formula in
Eq.~(\ref{eq:beforeScalarProduct}):%
\begin{equation}\label{eq:toThreeJSpecial}
\left(
\begin{array}{ccc}
a & b & i\\ 
1 & -1 & 0
\end{array}
\right)
=
\left(
\begin{array}{ccc}
a & b & i\\ 
0 & 0 & 0
\end{array}
\right)
\frac{{\mathcal J}^2_{i} - {\mathcal J}^2_{a} - {\mathcal J}^2_{b}}
{2 \sqrt{{\mathcal J}^2_{a}{\mathcal J}^2_{b}}},
\end{equation}
provided that $a+b+i$ is even~\cite{brink93Special}. On the
other hand, if $a+b+i$ is odd, then Eq.~(\ref{eq:toThreeJSpecial}) does not
hold; however, the second $3j$-coefficient of Eq.~(\ref{eq:beforeScalarProduct})
is zero in this case, so that we can safely plug Eq.~(\ref{eq:toThreeJSpecial})
into Eq.~(\ref{eq:beforeScalarProduct}):
\begin{eqnarray}
\lefteqn{
(1-\xi^2)
P'_{a}(\xi)P'_{b}(\xi)
}
\quad &
\nonumber
\\
&=&
 \sum_{i=|a-b|}^{a+b}
\frac{2i+1}{2}
({\mathcal J}^2_{a} + {\mathcal J}^2_{b} - {\mathcal J}^2_{i})
\left(
\begin{array}{ccc}
a & b & i\\ 
0 & 0 & 0
\end{array}
\right)
^2
P_{i}(\xi).
\label{eq:decompProductDerivLeg}
\end{eqnarray}
This simple expansion over the Legendre polynomials is of particular
interest in the sequel and will be used many times.

\subsubsection{Final steps}

With the help of Eq.~(\ref{eq:decompProductDerivLeg}), the integration of
Eq.~(\ref{eq:fundamentalIntegration2}) is straightforward if we know the
coefficients of $P'_{c}(\xi)$ over the Legendre polynomials; they
are easily obtained through an integration by parts:
\begin{equation}\label{eq:expandLegendreDer}
\int_{-1}^{1}\!
  d\xi \, P_{i}(\xi) P'_{c}(\xi)
=
\left\{
\begin{array}{ll}
2 & \mbox{ if $i+c$ odd and $i<c$}\\
0 & \mbox{ otherwise.}\\
\end{array}
\right. 
\end{equation}
Taking into account Eqs.~(\ref{eq:decompProductDerivLeg}) and
(\ref{eq:expandLegendreDer}), we thus arrive at our final formula:
\widetext
\begin{equation}\label{eq:fundamentalIntegration2:final}
\int_{-1}^{1}\!
  d\xi \,
(1-\xi^2)
P'_{a}(\xi)P'_{b}(\xi)P'_{c}(\xi)
=
\left\{
\begin{array}{l}
\displaystyle
\sum_{
\scriptsize
\parbox{20ex}{
\centering

$i=|a-b|$\\
$i+|a-b|$ even

}}^{c-1}
\left(
\begin{array}{ccc}
a & b & i\\ 
0 & 0 & 0
\end{array}
\right)
^2(2i+1)
 ({\mathcal J}^2_{a} + {\mathcal J}^2_{b} - {\mathcal J}^2_{i})
\quad
\mbox{if $a+b+c$ odd}
\\
\\
0
\quad
\mbox{otherwise.}
\end{array}
\right.
\end{equation}
\narrowtext\noindent
which is contained in Eq.~(\ref{eq:resultLegendre}) and
(\ref{eq:fundamentalIntegration2}).

The angular coefficient $A^{11}_{\kappa,\kappa_n}$ obtained in Eq.~(\ref{eq:a11:1}) can
then directly be evaluated with the help of results
(\ref{eq:fundamentalIntegration}) and
(\ref{eq:fundamentalIntegration2:final}), and we directly obtain our
final expression (\ref{eq:a11final}) for this coefficient.

 The evaluation of $A^{11}_{\kappa,\kappa_n}$ that we presented in this
  section also shows that only  terms of Eq.~(\ref{eq:a11general}) with $l$
  between $|l_{\kappa} - l_{-\kappa_n}|$ and
  $l_{\kappa} + l_{-\kappa_n}$ are non-zero; this fact,
  which was not obvious in the original formula, is explicitly
  expressed in our final result~(\ref{eq:a11final}).  

\subsection{Integration in $A^{12}_{\kappa,\kappa_n}$}
\label{sec:integA12}

The angular coefficient $A^{12}_{\kappa,\kappa_n}$ is more difficult to evaluate
than $A^{11}_{\kappa,\kappa_n}$, in particular because second-order derivatives of
$T$ appear in Eq.~(\ref{eq:a12general}). We show in this section how we
calculate the integration over $\xi$ in Eq.~(\ref{eq:a12general}) and obtain
the final expression Eq.~(\ref{eq:a12final}) [which must be combined with
Eq.~(\ref{eq:lMinTerm}) in numerical applications].

A first step consists in unifying the various terms of $A^{12}_{\kappa,\kappa_n}$
in Eq.~(\ref{eq:a12general}); we can replace the
$P_{l_{\kappa_n-1}}(\xi)=P_{l_{-\kappa_n}}(\xi)$ of the last
term of Eq.~(\ref{eq:a12general}) by $P_{l_{\kappa_n}}(\xi)$ and
$P'_{l_{\kappa_n}}(\xi)$ with the help
of
\begin{equation} %
P_{l_{-\kappa_n}}(\xi)
=
\frac{1}{\kappa_n} P'_{l_{\kappa_n}}(\xi)
+
\xi P_{l_{\kappa_n}}(\xi),
\end{equation}
which can be easily deduced from~\cite{edm96}, and
which implies that
\begin{eqnarray*} %
\lefteqn{
x_2 P_{l_{\kappa_n}}(\xi) - x_1 P_{l_{-\kappa_n}}(\xi)
}
\quad &
\\
&=&
(x_2 - x_1\xi) P_{l_{\kappa_n}}(\xi) - \frac{x_1}{\kappa_n} (1-\xi^2) P'_{l_{\kappa_n}}(\xi).
\end{eqnarray*}
The complicated factor $x_2-x_1 \xi$ can be removed with
\begin{equation} %
x_2 - x_1\xi = \rho \frac{\partial}{\partial x_2} \rho,
\end{equation}
which comes directly from the definition of $\rho$ in Eq.~(\ref{eq:defRho}).
By using $\partial / \partial x_2 = (\partial \rho / \partial x_2)
d / d \rho$, we finally obtain a special
expression for the last term of $A^{12}_{\kappa,\kappa_n}$ in Eq.~(\ref{eq:a12general})
\begin{eqnarray}
\lefteqn{
[x_2 P_{l_{\kappa_n}}(\xi) - x_1 P_{l_{\kappa_n-1}}(\xi)]
\frac{1}{\rho} \frac{d}{d\rho}
}
\quad &
\nonumber
\\
&=&
P_{l_{\kappa_n}}(\xi) \frac{\partial}{\partial x_2}
-
\frac{x_1}{\kappa_n} (1-\xi^2) P'_{l_{\kappa_n}}(\xi)
\frac{1}{\rho} \frac{d}{d\rho},
\label{eq:minusKNremoved}
\end{eqnarray}
where $P_{l_{\kappa_n-1}}(\xi)$ appears only in the left-hand side.
Equation (\ref{eq:minusKNremoved}) allows us to mix the various terms
of the original $A^{12}_{\kappa,\kappa_n}$ of Eq.~(\ref{eq:a12general}) in a uniformed
expression from which $P_{l_{\kappa_n-1}}(\xi)$ has disappeared:
\widetext
\begin{eqnarray}
\nonumber
A^{12}_{\kappa,\kappa_n}
& = &
-|\kappa|
\int_{-1}^{1}\!
  d\xi \,
\bigg(
\frac{\partial}{\partial x_2}
\left\{
[ x_2 P_{l_{\kappa-1}}(\xi) - x_1 P_{l_{\kappa}}(\xi) ]
P_{l_{\kappa_n}}(\xi)
\frac{1}{\rho}\frac{d}{d\rho} T(\rho)
\right\}
-
\frac{1}{\kappa\kappa_n}(1-\xi^2)P'_{l_{\kappa-1}}(\xi)
P'_{l_{\kappa_n}}(\xi)
\frac{1}{\rho}\frac{d}{d\rho} T(\rho)
\\
\label{eq:a12minusKNremoved}
&&\qquad\qquad\quad
+[x_2 P_{l_{\kappa-1}}(\xi)-x_1 P_{l_{\kappa}}(\xi)]
\frac{-x_1}{\kappa_n}(1-\xi^2) P'_{l_{\kappa_n}}(\xi)
{\left(\frac{1}{\rho}\frac{d}{d\rho}\right)}^2 T(\rho)
\bigg).
\end{eqnarray}

It is useful to do a similar operation in which
$P_{l_{\kappa-1}}(\xi)$ is replaced by $P_{l_{\kappa}}(\xi)$
and $P'_{l_{\kappa}}(\xi)$, so that the terms are more uniform:
by using again Eq.~(\ref{eq:minusKNremoved}), but with $\kappa$ instead of $\kappa_n$, we
obtain
\begin{eqnarray} %
\nonumber
A^{12}_{\kappa,\kappa_n}
& = &
-|\kappa|
\int_{-1}^{1}\!
  d\xi \,
\Bigg(
\frac{\partial}{\partial x_2}
\left\{
P_{l_{\kappa_n}}(\xi)
\left[
x_2 \frac{1}{\kappa} (1-\xi^2) P'_{l_{\kappa}}(\xi)
\frac{1}{\rho}\frac{d}{d\rho} T
-
P_{l_{\kappa}}(\xi) \frac{\partial}{\partial x_1} T
\right]
\right\}
\\
\nonumber
&&\qquad\qquad\quad
-
\frac{1}{\kappa\kappa_n}(1-\xi^2)P'_{l_{\kappa-1}}(\xi)
P'_{l_{\kappa_n}}(\xi)
\frac{1}{\rho}\frac{d}{d\rho} T
\\
\label{eq:a12:mKn:and:mK:removed}
&&\qquad\qquad\quad
+\frac{-x_1}{\kappa_n}(1-\xi^2) P'_{l_{\kappa_n}}(\xi)
\left[
x_2 \frac{1}{\kappa} (1-\xi^2)P'_{l_{\kappa}}(\xi)
{\left(\frac{1}{\rho}\frac{d}{d\rho}\right)}^2 T
-
P_{l_{\kappa}}(\xi) \frac{\partial}{\partial x_1} \frac{1}{\rho}\frac{d}{d\rho} T
\right]
\Bigg).
\end{eqnarray}

As seen before, the derivatives $d / d\rho$ are
fruitfully calculated with Eq.~(\ref{eq:differentiationRho}), that we apply
everywhere possible in Eq.~(\ref{eq:a12:mKn:and:mK:removed}):
\begin{eqnarray} %
\nonumber
A^{12}_{\kappa,\kappa_n}
& = &
-|\kappa|
\int_{-1}^{1}\!
  d\xi \,
\Bigg(
\frac{\partial}{\partial x_2}
\left\{
P_{l_{\kappa_n}}(\xi)
\left[
 \frac{-1}{\kappa x_1} (1-\xi^2) P'_{l_{\kappa}}(\xi)
\frac{\partial}{\partial \xi} T
-
P_{l_{\kappa}}(\xi) \frac{\partial}{\partial x_1} T
\right]
\right\}
\\
\nonumber
&&\qquad\qquad\quad
-
\frac{1}{\kappa\kappa_n}
\frac{-1}{x_1 x_2}
(1-\xi^2)P'_{l_{\kappa-1}}(\xi)
P'_{l_{\kappa_n}}(\xi)
\frac{\partial}{\partial \xi} T
\\
\label{eq:a12onlyXi}
&&\qquad\qquad\quad
+\frac{1}{\kappa_n x_2}(1-\xi^2) P'_{l_{\kappa_n}}(\xi)
\left[
 \frac{-1}{\kappa x_1} (1-\xi^2)P'_{l_{\kappa}}(\xi)
\frac{\partial^2}{\partial \xi^2}  T
-
P_{l_{\kappa}}(\xi) \frac{\partial}{\partial x_1} \frac{\partial}{\partial \xi} T
\right]
\Bigg).
\end{eqnarray}
\narrowtext\noindent
We note, however, that a \emph{second}-order derivative
$\partial^2 T/ \partial\xi^2 $ appears in the first term of the last
line in Eq.~(\ref{eq:a12onlyXi}). In order to obtain a simple
expression in which only \emph{first}-order derivatives are
present, we can use the following sort of integration by parts with
Legendre polynomials:
\begin{eqnarray} %
\lefteqn{
\int_{-1}^{1}\!
  d\xi \, (1-\xi^2) P'_{l}(\xi) g(\xi) f'(\xi)
}
\quad &
\nonumber
\\
&=&
\int_{-1}^{1}\!
  d\xi \, l(l+1) P_{l}(\xi) g(\xi) f(\xi)
\nonumber
\\
&&{}-
\int_{-1}^{1}\!
  d\xi \, (1-\xi^2) P'_{l}(\xi) g'(\xi) f(\xi),
\label{eq:legendreIbP}
\end{eqnarray}
can easily be proved by integrating by parts and by using the
differential equation of the Legendre
polynomials~\cite{aas72}:
\begin{equation}\label{eq:legendreDiffEq}
\frac{d}{d\xi}\left[ (1-\xi^2) P'_{l}(\xi) \right ]
=
-l(l+1) P_{l}(\xi).
\end{equation}
We thus transform the first term on the last line of Eq.~(\ref{eq:a12onlyXi})
with~Eq.~(\ref{eq:legendreIbP}) [with $l\equiv l_{\kappa_n}$, $f\equiv \partial T/
\partial\xi$ and $g \equiv (1-\xi^2)P'_{l_{\kappa}}(\xi)$]:
this term contains
\widetext
\begin{equation} \label{eq:noScndOrderDeriv} %
\int_{-1}^{1}\!
  d\xi \,
(1-\xi^2) P'_{l_{\kappa_n}}(\xi)
(1-\xi^2)P'_{l_{\kappa}}(\xi)
\frac{\partial^2}{\partial \xi^2}  T
=
\int_{-1}^{1}\!
  d\xi \, (1-\xi^2)
[
{\mathcal J}^2_{\kappa_n} P_{l_{\kappa_n}}(\xi)P'_{l_{\kappa}}(\xi)
+ {\mathcal J}^2_{\kappa} P'_{l_{\kappa_n}}(\xi) P_{l_{\kappa}}(\xi)
]
\frac{\partial}{\partial \xi} T,
\end{equation}
\narrowtext\noindent
where we have used the differential equation (\ref{eq:legendreDiffEq})
on the $g$ function, as well as the simple algebraic identity
$l_{\kappa} (l_{\kappa} + 1) = \kappa (\kappa+1)$, \emph{i.e.},
${\mathcal J}^2_{l_{\kappa}} = {\mathcal J}^2_{\kappa}$. 

We next use the partial-wave expansion (\ref{eq:devT}) of the $T$
function in order to simplify integrations over angles; since $T$ depends
on $\xi$ only through Legendre polynomials [Eq.~(\ref{eq:devT})], inserting
Eq.~(\ref{eq:noScndOrderDeriv}) into the angular coefficient (\ref{eq:a12onlyXi})
yields only integrals of one of the following forms:
\begin{mathletters}
\label{eq:basicIntegrals}
\begin{eqnarray}
&&\int_{-1}^{1}\!
  d\xi \, P_{a}(\xi) P_{b}(\xi) P_{c}(\xi)\\
&&\int_{-1}^{1}\!
  d\xi \, (1-\xi^2) P'_{a}(\xi) P'_{b}(\xi) P_{c}(\xi)\\
&&\int_{-1}^{1}\!
  d\xi \, (1-\xi^2) P'_{a}(\xi) P'_{b}(\xi) P'_{c}(\xi),
\end{eqnarray}
\end{mathletters}
where the third Legendre polynomial of each line comes from Eq.~(\ref{eq:devT}).
We have already all the tools that allow us to calculate them:
Eqs.~(\ref{eq:fundamentalIntegration}), (\ref{eq:fundamentalIntegration2}) and
(\ref{eq:decompProductDerivLeg}), along with the orthogonality
relation~\cite{aas72}:
\begin{equation}\label{eq:scalarProdLeg}
\int_{-1}^{1}\!
  d\xi \, P_{a}(\xi) P_{b}(\xi)
= \frac{2\delta_{a,b}}{2 a + 1}.
\end{equation}
After using expansion~(\ref{eq:devT}) and the values for
integrals~(\ref{eq:basicIntegrals}), we arrive directly at the final
result~(\ref{eq:a12final}) for the coefficient $A^{12}_{\kappa,\kappa_n}$. We
insist on the fact that the $l=|l_{\kappa} - l_{\kappa_n}|$
term of Eq.~(\ref{eq:a12final}) must, however, be evaluated with Eq.~(\ref{eq:lMinTerm})
in numerical calculations.

As for $A^{11}_{\kappa,\kappa_n}$, the evaluation of $A^{12}_{\kappa,\kappa_n}$ made in
  this section shows that the sum in Eq.~(\ref{eq:a12general}) contains a
  \emph{finite} number of terms: only the terms of Eq.~(\ref{eq:a12general})
  with $l$ between $|l_{\kappa} - l_{\kappa_n}|$ and
  $l_{\kappa} + l_{\kappa_n}$ are non-zero, as expressed
  in our final result~(\ref{eq:a12final}).  

\section{Alternative method for the calculation of the angular coefficients}
\label{subsec:direct}

In order to check our final results (\ref{eq:a11final}),
(\ref{eq:a12final}) and (\ref{eq:lMinTerm}), we derived and
implemented independently a second set of expressions, that we
compared to the results of the method presented in Sec.~\ref{sec:optnumder}
over a wide range of arguments. This allowed us to limit the number of
comparisons with \emph{Mathematica}; in fact, direct integrations of
Eqs.~(\ref{eq:a11general}) and (\ref{eq:a12general}) lead to very lengthy
calculations.  We found that the method presented in this subsection
is accurate almost everywhere, but that it is slower than the method
presented in Sec.~\ref{sec:form-angul} and \ref{sec:optnumder}.  We present
here the main steps of an alternate calculation of the integrals of
Eqs.~(\ref{eq:a11general}) and (\ref{eq:a12general}).

\subsection{Angular integration in $A^{11}_{\kappa,\kappa_n}$}

We evaluate $A^{11}_{\kappa,\kappa_n}$, starting again from the most detailed
basic expression, Eq.~(\ref{eq:a11general}). The idea of this second evaluation
is to express the integrand \emph{solely} in terms of Legendre
polynomials of the variable $\xi$; doing so will  allow us to use
the integration result of Eq.~(\ref{eq:fundamentalIntegration}).

With this goal in mind, we remove factors $1-\xi^2$ and $\xi$ in
Eq.~(\ref{eq:a11general}) with the following Legendre polynomial
identity [adapted to our particular form (\ref{eq:defOrbital}) of orbital
angular momenta]:
\begin{equation}
\frac{1}{\kappa}\left(1-\xi^2\right)
P'_{l_{\kappa}}(\xi)=P_{l_{-\kappa}}(\xi)-\xi P_{l_{\kappa}}(\xi),
\label{eq:lepolid}
\end{equation}
and with the help of the Legendre recursion relation
\begin{equation}
\xi P_l(\xi)=\frac{1}{2l+1}\left[ (l+1) P_{l+1}(\xi)+lP_{l-1}(\xi)\right].
\label{eq:lepolrec}
\end{equation}
The $T$ function of Eq.~(\ref{eq:devT}) can then be expanded with Legendre
polynomials with Eqs.~(\ref{eq:devT}) and (\ref{eq:differentiationRho}), and we
can integrate once with Eq.~(\ref{eq:fundamentalIntegration}); we thus get
\begin{eqnarray}
A^{11}_{\kappa,\kappa_n}&=& b |\kappa| \sum_{l \ge 0} (2l+1)j_{l}\left(y_2\right) j_{l}\left(y_1\right)\nonumber \\
&\times& \bigg\{
2  \left(
\begin{array}{ccc}
l_{\kappa}  & l_{-\kappa_n} & l \\ 0 & 0 & 0
\end{array}
\right)^2  
 \nonumber \\
&&+
\frac{1}{\kappa_n y_2 y_1 }
\int_{-1}^{1} d\xi \bigg[
P_{l_{-\kappa}}(\xi)
P'_{l_{-\kappa_n}}(\xi)  P'_{l}(\xi) \nonumber\\
&&-\frac{1}{ 2l_{\kappa} + 1}
\left(l_{\kappa}+1\right) P_{l_{\kappa}+1}(\xi)
P'_{l_{-\kappa_n}}(\xi)  P'_{l}(\xi) \nonumber \\
&&-  \frac{1}{ 2l_{\kappa} + 1}
l_{\kappa} P_{l_{\kappa}-1}(\xi)
P'_{l_{-\kappa_n}}(\xi) P'_{l}(\xi)
\bigg]\bigg\}.
\label{eq:a11d}
\end{eqnarray}

Formula (\ref{eq:a11d}) contains only \emph{one} type of integral, namely:
\begin{equation}\label{eq:defDone}
{\cal D}^{1:(n,l)}_{m} \equiv \int_{-1}^{1} d\xi  P_{m}(\xi) P'_{l}(\xi) P'_{n} (\xi).
\end{equation}
The integrand of this formula can be transformed into a linear
combination of products of three Legendre polynomials [easily
integrated with Eq.~(\ref{eq:fundamentalIntegration})], by expanding the
derivatives of Legendre polynomials over the orthogonal basis of
Legendre polynomials: in fact, Eqs.~(\ref{eq:scalarProdLeg}) and
(\ref{eq:expandLegendreDer}) yield
\begin{equation}\label{eq:derivativeLeg}
P'_{n}(\xi)=\sum_{m=0}^{n-1} (2 m+1) P_{m}(\xi)\delta_p(n,m),
\end{equation}
where
\[
\delta_p(n,m) \equiv \left\{
\begin{array}{ll}
0 & \mbox{if } n \mbox{ and }m\mbox{ have the same parity,}
\\
1 & \mbox{otherwise.}
\\
\end{array}
\right. 
\]
In summary, we obtain an evaluation of the quantity ${\cal
  D}^{1:(n,l)}_{m}$ of Eq.~(\ref{eq:defDone}) with the help of
Eqs.~(\ref{eq:derivativeLeg}) and (\ref{eq:fundamentalIntegration}):
\widetext
\begin{eqnarray}
{\cal D}^{1:(n,l)}_{m}&=&
 \int_{-1}^{1} d\xi  P_{m}(\xi)\sum_{i=0}^{l-1} (2 i+1) P_{i}(\xi)\delta_p(l,i)
\sum_{j=0}^{n-1} (2 j+1) P_{j}(\xi)\delta_p(n,j) \nonumber \\
&=& \sum_{i=0}^{l-1}\sum_{j=0}^{n-1}(2 i+1)(2 j+1)\delta_p(n,j)\delta_p(l,i)
\int_{-1}^{1} d\xi P_{m}(\xi)P_{i}(\xi)P_{j}(\xi)\nonumber \\
&=& 2\sum_{i=0}^{l-1}\sum_{j=0}^{n-1}(2 i+1)(2 j+1)\delta_p(l,i)\delta_p(n,j)
\left(
\begin{array}{ccc}
m& i & j \\ 0 & 0 & 0
\end{array}
\right)^2.
\label{eq:expandD}
\end{eqnarray}

With definition (\ref{eq:defDone}), the final result of this section
for $A^{11}_{\kappa,\kappa_n}$ can be deduced from Eq.~(\ref{eq:a11d}) and reads:
\begin{eqnarray}
A^{11}_{\kappa,\kappa_n}&=& b |\kappa| \sum_{l \ge 0} (2l+1)j_{l}\left(y_2\right)
j_{l}\left(y_1\right)\nonumber 
\\
&&\times 
\bigg\{
2  \left(
\begin{array}{ccc}
l_{\kappa}  & l_{-\kappa_n} & l \\ 0 & 0 & 0
\end{array}
\right)^2  
+\frac{1}{\kappa_n y_2 y_1 }\bigg[
{\cal D}^{1:(l_{-\kappa_n},\, l)}_{l_{-\kappa}}
-\frac{1}{ 2l_{\kappa} + 1}
\left(l_{\kappa}+1\right){\cal
  D}^{1:(l_{-\kappa_n},\, l)}_{l_{\kappa}+1}  
-  \frac{1}{ 2l_{\kappa} + 1}
 l_{\kappa} {\cal D}^{1:(
l_{-\kappa_n},\, l)}_{l_{\kappa}-1}
\bigg]\bigg\},
\label{eq:a11kp}
\end{eqnarray}
where the quantity ${\cal D}^{1:(n,l)}_{m}$ is given explicitly by
Eq.~(\ref{eq:expandD}).
\narrowtext

A numerical implementation of Eq.~(\ref{eq:expandD}) can make use of the fact
that parity properties of the Legendre polynomials
\cite{edm96} immediately impose
\[ 
{\cal D}^{1:(n,l)}_{m} = 0 \mbox{ if } l+m+n \mbox{ odd}.
\]
Moreover, numerical implementations are facilitated by the fact that
the summation over $l$ in Eq.~(\ref{eq:a11kp}) contains a \emph{finite} number
of non-zero terms---even though this is not obvious from the above
form---, as noted for definition (\ref{eq:a11general}) in
Sec.~\ref{sec:optnumder}: in Eq.~(\ref{eq:a11kp}), the summation over $l$ can be
restricted to the range $|l_{\kappa} - l_{-\kappa_n}|$
\ldots ${l_{\kappa} - l_{-\kappa_n}}$.

\subsection{Off-diagonal element}

To evaluate $A^{12}_{\kappa,\kappa_n}$ we start from the most general published
formula [Eq.~(\ref{eq:a12general})]. As for $A^{12}_{\kappa,\kappa_n}$, the idea
of the derivation to follow consists in transforming integrations over
$\xi$ into integrals of products of Legendre polynomials \emph{only},
so that we can use the integration result
(\ref{eq:fundamentalIntegration}). The first step is again
accomplished by removing $1-\xi^2$ factors with the help of
Eqs.~(\ref{eq:lepolid}) and (\ref{eq:lepolrec}), and by calculating derivatives
of $T$ with Eq.~(\ref{eq:differentiationRho}):
\widetext
\begin{eqnarray}
A^{12}_{\kappa,\kappa_n}%
&=&\frac{|\kappa|}{ x_2 x_1} \int_{-1}^{1} d\xi 
\bigg\{
P_{l_{-\kappa}}(\xi)P_{l_{\kappa_n}}(\xi) \frac{\partial }{\partial \xi}T(\rho)
- \frac{1}{\kappa_n }
P_{l_{\kappa}}(\xi)
P'_{l_{\kappa_n}}(\xi) 
 \frac{\partial }{\partial \xi} T(\rho)
 \nonumber \\
&& 
\quad 
{}+ \frac{1}{\kappa_n  (2  l_{-\kappa}+1)}
\left(l_{-\kappa}+1\right) P_{l_{-\kappa}+1}(\xi)
P'_{l_{\kappa_n}}(\xi) 
 \frac{\partial }{\partial \xi} T(\rho)
 + \frac{1}{\kappa_n (2 l_{-\kappa}+1)}
 l_{-\kappa} P_{l_{-\kappa}-1}(\xi)
P'_{l_{\kappa_n}}(\xi) 
 \frac{\partial }{\partial \xi} T(\rho)\nonumber \\
&&
\quad
{}-\frac{1}{x_2 x_1} \bigg[ x_2 P_{l_{-\kappa}}(\xi) -
 x_1 P_{l_{\kappa}}(\xi) \bigg]
\bigg[ x_2 P_{l_{\kappa_n}}(\xi) - 
x_1 P_{l_{-\kappa_n}}(\xi)  \bigg]
\frac{\partial^2 }{\partial \xi^2} T(\rho)
\bigg\},
\end{eqnarray}
where we have used the simple identity on orbital momenta
$l_{\kappa-1} = l_{-\kappa}$.
\narrowtext

A replacement of $T$ by its partial-wave expansion (\ref{eq:devT})
directly yields:
\widetext
\begin{eqnarray}
A^{12}_{\kappa,\kappa_n}
&=&
\frac{|\kappa|}{ b  x_2 x_1} \sum_{l=0}^\infty
(2l+1)j_{l}\left(y_1\right) j_{l}\left(y_2\right)
\int_{-1}^{1} d\xi 
\bigg\{
P_{l_{-\kappa}}(\xi)P_{l_{\kappa_n}}(\xi) P'_{l}(\xi)
- \frac{1}{\kappa_n }
P_{l_{\kappa}}(\xi)
P'_{l_{\kappa_n}}(\xi) 
 P'_{l}(\xi) \nonumber \\
&& + \frac{1}{\kappa_n (2l_{-\kappa}+1)}
\left(l_{-\kappa}+1\right) P_{l_{-\kappa}+1}(\xi)
P'_{l_{\kappa_n}}(\xi) 
 P'_{l}(\xi)
 + \frac{1}{\kappa_n (2l_{-\kappa}+1)}
\left( l_{-\kappa}\right)P_{l_{-\kappa}-1}(\xi)
P'_{l_{\kappa_n}}(\xi) 
 P'_{l}(\xi)\nonumber \\
&&-\bigg[\frac{x_2}{x_1}
 P_{l_{-\kappa}}(\xi) P_{l_{\kappa_n}}(\xi)P''_{l}(\xi)
- P_{l_{-\kappa}}(\xi) P_{l_{-\kappa_n}}(\xi)P''_{l}(\xi)
-  P_{l_{\kappa}}(\xi) P_{l_{\kappa_n}}(\xi) P''_{l}(\xi)
+ \frac{x_1}{x_2 } P_{l_{\kappa}}(\xi) P_{l_{-\kappa_n}}(\xi) P''_{l}(\xi)
\bigg]\bigg\}.
\end{eqnarray}
\narrowtext

There are only two new categories of terms to evaluate. We thus define 
\begin{equation}\label{eq:firstD}
{\cal D}^{1:l}_{m,n}=\int_{-1}^{1} d\xi  P_{m}(\xi) P_{n}(\xi)P'_{l}(\xi)
\end{equation}
and
\begin{equation}\label{eq:secondD}
{\cal D}^{2:l}_{m,n}=\int_{-1}^{1} d\xi  P_{m} (\xi)P_{n}(\xi)P''_{l}(\xi).
\end{equation}
Both these quantities can be expressed by means of an expansion of the
Legendre polynomials derivatives with Eq.~(\ref{eq:derivativeLeg}). We thus
evaluate the first expression as
\begin{eqnarray}
{\cal D}^{1:l}_{m,n}&=&
 \int_{-1}^{1} d\xi  P_{m}(\xi)\sum_{i=0}^{l-1} (2 i+1) P_{i}(\xi)\delta_p(l,i)
 P_{n}(\xi)\nonumber \\
&=&2\sum_{i=0}^{l-1} (2 i+1)\delta_p(l,i)
\left(
\begin{array}{ccc}
m& i & n \\ 0 & 0 & 0
\end{array}
\right)^2,
\label{eq:expandDtwo}
\end{eqnarray}
where we have integrated products of three Legendre polynomials with
(\ref{eq:fundamentalIntegration}. This formula is coherent with the
fact that symmetry properties of Legendre
polynomials~\cite{edm96} show that expression
(\ref{eq:firstD}) yields $0$ if $m+n+l$ is even; this property should
be used in numerical implementations.

The function  with a second-order derivative [Eq.~(\ref{eq:secondD})] can be
evaluated by using an integration by parts:
\begin{eqnarray}
{\cal D}^{2:l}_{m,n}&=&\int_{-1}^{1} d\xi  P_{m}(\xi) P_{n}(\xi)P''_{l}(\xi) \nonumber \\
&=&\left[ P_{m}(\xi) P_{n}(\xi)P'_{l}(\xi)\right]^1_{-1}
\nonumber\\
&&
\quad
{}
-\int_{-1}^{1}
d\xi  \left[P_{n}(\xi)P_{m}(\xi)\right]' P'_{l}(\xi)\nonumber \\
&=&\left[ P_{m}(\xi)P_{n}(\xi) P'_{l}(\xi)\right]^1_{-1}-\int_{-1}^{1} d\xi
  P_{n}(\xi)P'_{m}(\xi) P'_{l}(\xi)
\nonumber\\
&&
\quad
{}
-\int_{-1}^{1} d\xi P_{m}(\xi)
  P_{n}'(\xi) P'_{l}(\xi)
\label{eq:expandDthree}
\end{eqnarray}
Using $P_{m}(1)=1$, $P_{m}(-1)=(-1)^m$, $P'_{l}(1)=l(l+1)/2$ and
$P'_{l}(-1)=-(-1)^l[l(l+1)/2]$, we get
\begin{eqnarray}
{\cal D}^{2:l}_{m,n}&=&
\frac{l(l+1)}{2}\left[1+(-1)^{m+l+n}\right]-\left({\cal D}^{1:(m,\, l)}_{ln}+{\cal D}^{1:(n,l)}_{m} \right).
\end{eqnarray}
Symmetry properties of the Legendre polynomials and definition
(\ref{eq:secondD}) show that this quantity is zero whenever $l+m+n$ is
odd, a fact that can be used in numerical calculations.

Our final expression for the angular coefficient $A^{12}_{\kappa,\kappa_n}$ thus reads:
\widetext
\begin{eqnarray}
A^{12}_{\kappa,\kappa_n}&=&\frac{|\kappa|}{ b  x_2 x_1} \sum_{l=0}^\infty
(2l+1)j_{l}\left(y_1\right) j_{l}\left(y_2\right)
\bigg[{\cal D}^{1:l}_{l_{-\kappa},\, l_{\kappa_n}}
- \frac{1}{\kappa_n }
{\cal D}^{1:(l_{\kappa_n},\, l)}_{l_{\kappa}}
+ \frac{1}{\kappa_n (2  l_{-\kappa}+1)}
\left(l_{-\kappa}+1\right) {\cal D}^{1:(l_{\kappa_n},\, l)}_{l_{-\kappa}+1}
\nonumber \\
&& 
\quad
{}
+ \frac{1}{\kappa_n (2  l_{-\kappa}+1)}
 l_{-\kappa} {\cal D}^{1:(l_{\kappa_n},\, l)}_{l_{-\kappa}-1}
-\frac{x_2}{x_1}
 {\cal D}^{2:l}_{l_{-\kappa},\, l_{\kappa_n}}
+ {\cal D}^{2:l}_{l_{-\kappa},\, l_{-\kappa_n}}
+ {\cal D}^{2:l}_{l_{\kappa},\, l_{\kappa_n}}
- \frac{x_1}{x_2 } {\cal D}^{2:l}_{l_{\kappa},\, l_{-\kappa_n}}
\bigg],
\label{eq:a12paulFinal}
\end{eqnarray}
\narrowtext\noindent
where the various ${\mathcal D}$ quantities are given explicitly in
Eqs.~(\ref{eq:expandD}), (\ref{eq:expandDtwo}) and (\ref{eq:expandDthree}). We
have checked the above formula by comparing Fortran and
\emph{Mathematica} outputs for many values relevant to the present
work. Furthermore, numerical implementations should make use of the
fact that it is sufficient to do the summation over $l$ in
Eq.~(\ref{eq:a12paulFinal})  only with $l$
in the range $|l_{\kappa} - l_{\kappa_n}|$ \ldots
$(l_{\kappa} + l_{\kappa_n})$, because all other terms are
zero [since Eq.~(\ref{eq:a12general}) contains only these terms, as
noted in Sec.~\ref{sec:optnumder}].

\section{Conclusion}
\label{sec:concl}

We have obtained in Eqs.~(\ref{eq:a11final}), (\ref{eq:a12final}) and
(\ref{eq:lMinTerm}) analytic formulas for angular coefficients that
appear in an efficient numerical method of calculation of the electron
self-energy in hydrogenlike atoms~\cite{moh74a}; only results for
electrons with angular momentum $1/2 \le j \le 3/2$ were previously
obtainable with this method~\cite{moh74a,mak92}. Recently developed
numerical renormalization
techniques~\cite{iam92,iam98,iam01} allowed us
to give numerical self-energy shifts for high-$j$ states ($5/2 \le j
\le 9/2$) of $n=3$, $4$ and $5$ levels in the range $60 \le Z \le 110$;
our results are in agreement with recently available
results~\cite{v99}, but we provide values for many other
nuclear charges $Z$, and the precision of our calculations is generally
greater by about two orders of magnitude. These numerical results
could for instance serve to include QED effects in many-body atomic
calculations in atoms and ions with 
electrons of high angular momentum~\cite{wcc2001}.

\acknowledgments

We are grateful to the CINES (Montpellier, France) for a grant of time
on its SP2 and SP3 parallel computers. We wish to thank Dr. U.
Jentschura for very interesting discussions.

\newpage

\narrowtext

\begin{table}
\caption{\label{tbl:firstTable}Scaled self-energy $F(Z\alpha)$ for
  $d_{5/2}$ states ($\kappa=-3$).}
\begin{tabular}{llll}
\multicolumn{1}{c}{$Z$} & \multicolumn{1}{c}{$n = 3$} & \multicolumn{1}{c}{$n = 4$} & \multicolumn{1}{c}{$n = 5$} \\
\hline
60 & 0.0503484(5) & 0.0540821(4) & 0.0560084(4) \\
66 & 0.0522327(4) & 0.0563571(4) & 0.0584608(5) \\
70 & 0.0535711(5) & 0.0579826(4) & 0.0602164(4) \\
74 & 0.0549734(4) & 0.0596935(4) & 0.0620673(3) \\
80 & 0.0571916(5) & 0.0624150(5) & 0.0650174(4) \\
83 & 0.0583498(4) & 0.0638430(4) & 0.0665677(4) \\
90 & 0.0611703(5) & 0.0673377(4) & 0.0703682(4) \\
92 & 0.0620040(5) & 0.0683750(5) & 0.0714981(9) \\
100 & 0.0654457(5) & 0.0726763(5) & 0.0761897(5) \\
110 & 0.0699331(6) & 0.0783217(5) & 0.0823592(6) \\
\end{tabular}
\end{table}

\begin{table}
\caption{Scaled self-energy $F(Z\alpha)$ for $d_{5/2}$ states ($\kappa=3$).}
\begin{tabular}{lll}
\multicolumn{1}{c}{$Z$} & \multicolumn{1}{c}{$n = 4$} & \multicolumn{1}{c}{$n = 5$} \\
\hline
60 & -0.0203415(3) & -0.0195626(3) \\
66 & -0.0201262(4) & -0.0193062(4) \\
70 & -0.0199699(4) & -0.0191179(4) \\
74 & -0.0198020(3) & -0.0189139(3) \\
80 & -0.0195284(4) & -0.0185761(3) \\
83 & -0.0193804(4) & -0.0183914(4) \\
90 & -0.0190034(4) & -0.0179156(3) \\
92 & -0.0188870(4) & -0.0177669(4) \\
100 & -0.0183786(4) & -0.0171113(4) \\
110 & -0.0176375(5) & -0.0161389(5) \\
\end{tabular}
\end{table}

\begin{table}
\caption{Scaled self-energy $F(Z\alpha)$ for $f_{7/2}$ states ($\kappa=-4$).}
\begin{tabular}{lll}
\multicolumn{1}{c}{$Z$} & \multicolumn{1}{c}{$n = 4$} & \multicolumn{1}{c}{$n = 5$} \\
\hline
60 & 0.0221590(3) & 0.0230942(5) \\
66 & 0.0225316(4) & 0.0235345(4) \\
70 & 0.0227973(3) & 0.0238498(3) \\
74 & 0.0230768(4) & 0.0241830(3) \\
80 & 0.0235224(4) & 0.0247170(3) \\
83 & 0.0237572(4) & 0.0249997(4) \\
90 & 0.0243367(4) & 0.0257006(3) \\
92 & 0.0245106(4) & 0.0259119(3) \\
100 & 0.0252427(4) & 0.0268060(4) \\
110 & 0.0262422(5) & 0.0280366(4) \\
\end{tabular}
\end{table}

\begin{table}
\caption{\label{tbl:lastTable}Scaled self-energy $F(Z\alpha)$ for $g_{9/2}$
  ($\kappa=-5$) and $g_{7/2}$ ($\kappa=4$) states.}
\begin{tabular}{lll}
\multicolumn{1}{c}{$Z$} & \multicolumn{1}{c}{$\kappa = -5$} & \multicolumn{1}{c}{$\kappa=4$}\\
\hline
60 & 0.0127593(2) & -0.0125408(4) \\
66 & 0.0128741(3)  & -0.0124866(3) \\
70 & 0.0129555(3)  & -0.0124484(4) \\
74 & 0.0130416(3) & -0.0124084(3)\\
80 & 0.0131782(3) & -0.0123450(4)  \\
83 & 0.0132500(3) & -0.0123115(3)  \\
90 & 0.0134266(3) & -0.0122290(3) \\
92 & 0.0134794(4) & -0.0122043(3)  \\
100 & 0.0137014(3)& -0.0120991(3)  \\
110 & 0.0140037(3) & -0.0119529(3) \\
\end{tabular}
\end{table}

\mediumtext

\begin{table}
\caption{\label{tbl:compYerokhinFirst}Scaled self-energy
    $F(Z\alpha)$ compared between previously published results and
    this work, for
    levels with principal quantum numbers $n=3$ and $4$. There is only
    one (small) incoherence between the results ($3d_{5/2}$ for
    $Z=74$).
}
\begin{tabular}{llllll}
\multicolumn{2}{c}{} & \multicolumn{1}{c}{$n=3$} & \multicolumn{3}{c}{$n=4$} \\
$Z$ && \multicolumn{1}{c}{$3d_{5/2}$} & \multicolumn{1}{c}{$4d_{5/2}$} & \multicolumn{1}{c}{$4f_{5/2}$} & \multicolumn{1}{c}{$4f_{7/2}$} \\
\hline
74 & Ref.~\cite{v99} & 0.0550(0) & 0.0598(4) & -0.0198(4) & 0.0231(4) \\
 & This work & 0.0549734(4) & 0.0596935(4) & -0.0198020(3) & 0.0230768(4) \\
83 & Ref.~\cite{v99} & 0.0583(0) & 0.0639(3) & -0.0194(3) & 0.0238(3) \\
 & This work & 0.0583498(4) & 0.0638430(4) & -0.0193804(4) & 0.0237572(4) \\
92 & Ref.~\cite{v99} & 0.0620(0) & 0.0684(2) & -0.0189(2) & 0.0245(2) \\
 & This work & 0.0620040(5) & 0.0683750(5) & -0.0188870(4) & 0.0245106(4) \\
\end{tabular}
\end{table}

\begin{table}\caption{\label{tbl:compYerokhinLast}Scaled self-energy
    $F(Z\alpha)$ compared between previously published results
    and this work, for
    levels with principal quantum number $n=5$. There are only two (small)
    incoherences: $5d_{5/2}$ for $Z=83$ and $5d_{5/2}$ for $Z=74$.
}
\begin{tabular}{lllllll}
\multicolumn{2}{c}{} & \multicolumn{5}{c}{$n=5$} \\
$Z$ && \multicolumn{1}{c}{$5d_{5/2}$} & \multicolumn{1}{c}{$5f_{5/2}$} & \multicolumn{1}{c}{$5f_{7/2}$} & \multicolumn{1}{c}{$5g_{7/2}$} & \multicolumn{1}{c}{$5g_{9/2}$} \\
\hline
74 & Ref.~\cite{v99} & 0.0628(6) & -0.0184(7) & 0.0247(7) & -0.0121(9) & 0.0135(9) \\
 & This work & 0.0620673(3) & -0.0189139(3) & 0.0241830(3) & -0.0124084(3) & 0.0130416(3) \\
83 & Ref.~\cite{v99} & 0.0671(5) & -0.0180(5) & 0.0254(6) & -0.0121(8) & 0.0136(8) \\
 & This work & 0.0665677(4) & -0.0183914(4) & 0.0249997(4) & -0.0123115(3) & 0.0132500(3) \\
92 & Ref.~\cite{v99} & 0.0719(5) & -0.0175(5) & 0.0262(5) & -0.0121(5) & 0.0136(6) \\
 & This work & 0.0714981(9) & -0.0177669(4) & 0.0259119(3) & -0.0122043(3) & 0.0134794(4) \\
\end{tabular}
\end{table}

\narrowtext
\newpage

\begin{figure}
\psfig{figure=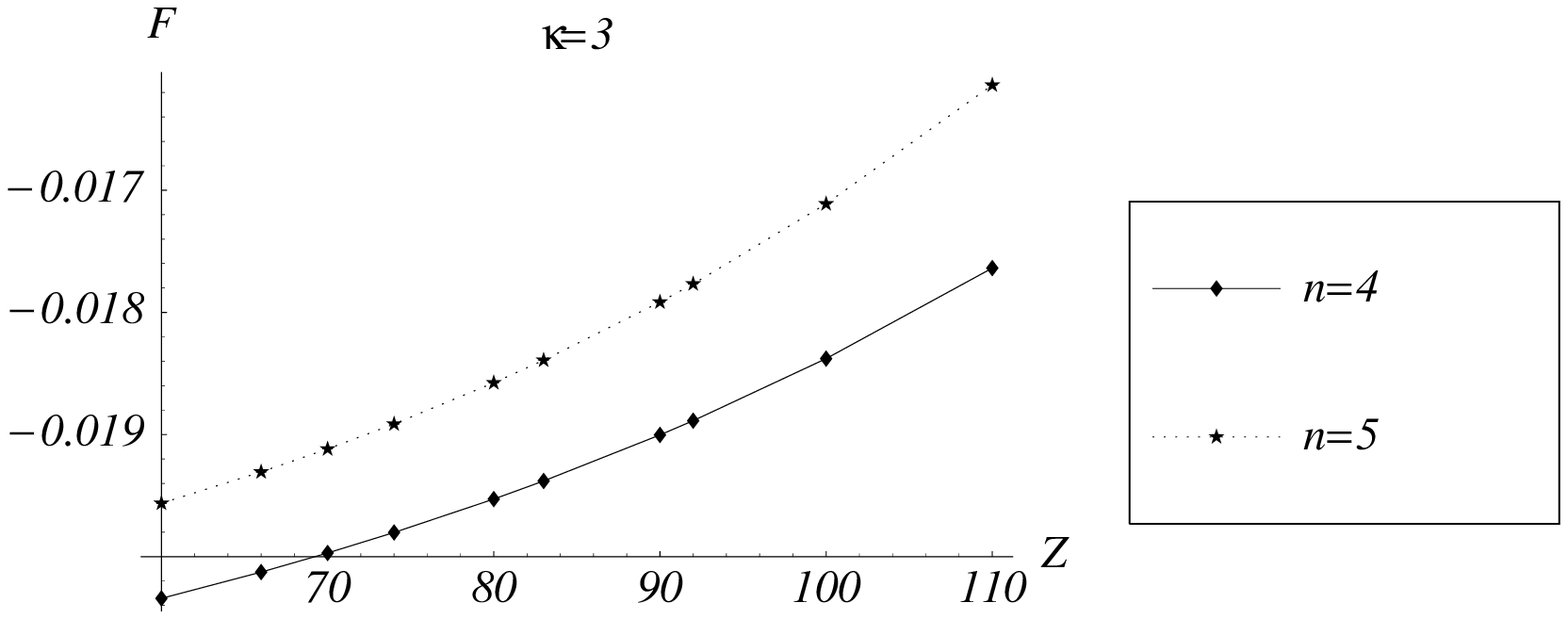}
\caption{Scaled self-energy $F(Z\alpha)$ for $f_{5/2}$ states of levels $n=4$
  and 5.}
\end{figure}

\end{document}